\catcode`\@=11

\newif\if@fewtab\@fewtabtrue

{\count255=\time\divide\count255 by 60
\xdef\hourmin{\number\count255}
\multiply\count255 by-60\advance\count255 by\time
\xdef\hourmin{\hourmin:\ifnum\count255<10 0\fi\the\count255}}
\def\ps@draft{\let\@mkboth\@gobbletwo
    \def\@oddhead{}
    \def\@oddfoot
       {\hbox to 7 cm{$\scriptstyle Draft\ version:\ \draftdate$
       \hfil}\hskip -7cm\hfil\rm\thepage \hfil}
    \def\@evenhead{}\let\@evenfoot\@oddfoot}

\def\ceqno{\global\@fewtabfalse
    \ifcase\@eqcnt \def\@tempa{& & &}\or \def\@tempa{& &}
      \or \def\@tempa{&}
      \or\def\@tempa{}\fi\@tempa
{\rm(\theequation)}}

\def\aeqno#1{\global\@fewtabfalse
    \ifcase\@eqcnt \def\@tempa{& & &}\or \def\@tempa{& &}
      \or \def\@tempa{&}
      \or\def\@tempa{}\fi\@tempa
{\rm(\theequation,#1)}}

\def\label#1{\ifnum\draftcontrol=1
 \global\def\draftnote{$\scriptstyle #1$}\fi
 \@bsphack\if@filesw {\let\thepage\relax
   \def\protect{\noexpand\noexpand\noexpand}%
\xdef\@gtempa{\write\@auxout{\string
      \newlabel{#1}{{\@currentlabel}{\thepage}}}}}\@gtempa
   \if@nobreak \ifvmode\nobreak\fi\fi\fi
  \@esphack}

\def\alabel#1#2{\label{#1}\global\@fewtabfalse
    \ifcase\@eqcnt \def\@tempa{& & &}\or \def\@tempa{& &}
      \or \def\@tempa{&}
      \or\def\@tempa{}\fi\@tempa
{\hbox to 3cm{\phantom{\rm(\theequation,#2)}
\draftnote \hfil}\hskip -3cm {\rm(\theequation,#2)}}}

\def\clabel#1{\label{#1}\global\@fewtabfalse
    \ifcase\@eqcnt \def\@tempa{& & &}\or \def\@tempa{& &}
      \or \def\@tempa{&}
      \or\def\@tempa{}\fi\@tempa
{\hbox to 3cm{\phantom{\rm(\theequation)}
\draftnote \hfil}\hskip -3cm{\rm(\theequation)}}}

\def\eqnarray{\def\draftnote{{}}\global\@fewtabtrue
\stepcounter{equation}\let\@currentlabel=\theequation
\global\@eqnswtrue
\global\@eqcnt\z@\tabskip\@centering\let\\=\@eqncr
$$\halign to \displaywidth\bgroup\@eqnsel\hskip\@centering\@eqcnt\z@
  $\displaystyle\tabskip\z@{##}$&\global\@eqcnt\@ne
  \hskip 1\arraycolsep \hfil${##}$\hfil
  &\global\@eqcnt\tw@ \hskip 1\arraycolsep
$\displaystyle\tabskip\z@{##}$
\hfil  \tabskip\@centering&\global\@eqcnt\thr@@\llap{##}\tabskip\z@
\cr}

\def\endeqnarray{\@@eqncr\egroup
      \global\advance\c@equation\m@ne$$\global\@ignoretrue}

\def\@eqnnum{\hbox to 3cm{\phantom{\rm(\theequation)} \draftnote
                         \hfil}\hskip -3cm {\rm(\theequation)}}

\def\@@eqncr{\let\@tempa\relax
    \ifcase\@eqcnt \def\@tempa{& & &}\or \def\@tempa{& &}
      \or \def\@tempa{&}
      \or\def\@tempa{}
\fi\@tempa
\if@eqnsw
\if@fewtab\@eqnnum\fi
\stepcounter{equation}\fi\global
\@eqnswtrue\global\@eqcnt\z@\global\@fewtabtrue\cr}

\def\draftcite#1{\ifnum\draftcontrol=1#1\else{}\fi}

\def\@lbibitem[#1]#2{\item{}\hskip -3cm \hbox to 2cm
{\hfil$\scriptstyle\draftcite{#2}$}\hskip
1cm[\@biblabel{#1}]\if@filesw
     {\def\protect##1{\string ##1\space}\immediate
      \write\@auxout{\string\bibcite{#2}{#1}}}\fi\ignorespaces}

\def\@bibitem#1{\item\hskip -3cm \hbox to 2cm
{\hfil $\scriptstyle\draftcite{#1}$}\hskip 1cm
\if@filesw \immediate\write\@auxout
       {\string\bibcite{#1}{\the\value{\@listctr}}}\fi\ignorespaces}

\font\tendl=msbm10  scaled \magstep1
\font\sevendl=msbm7 scaled \magstep1
\font\fivedl=msbm5 scaled \magstep1
\font\tengl=eufm10  scaled \magstep1
\font\sevengl=eufm7 scaled \magstep1
\font\fivegl=eufm5 scaled \magstep1

\newfam\dlfam  
\textfont\dlfam=\tendl \scriptfont\dlfam=\sevendl
\scriptscriptfont\dlfam=\fivedl
\newfam\glfam  
\textfont\glfam=\tengl \scriptfont\glfam=\sevengl
\scriptscriptfont\glfam=\fivegl

\def\draftdate{\number\month/\number\day/\number\year\ \ \ \hourmin }

\global\def\draftcontrol{0}
\catcode`\@=12
\def\tilde{\widetilde}

\documentstyle[12pt,epsf]{article}
\def\theequation{{\arabic{equation}}}
\newcommand{\be}{\begin{eqnarray}}
\newcommand{\en}{\end{eqnarray}\vs 0.5 cm}
\newcommand{\non}{\nonumber}
\newcommand{\no}{\noindent}
\newcommand{\vs}{\vskip}
\newcommand{\hs}{\hspace}

\newcommand{\un}{\underline}
\newcommand{\lsim}{{\smash{\mathop{<}\limits_{^\sim}}}}
\newcommand{\gsim}{{\smash{\mathop{>}\limits_{^\sim}}}}
\newcommand{\Nom}{{\bf\omega}}
\newcommand{\Na}{{\bf a}}
\newcommand{\Nu}{{\bf u}}
\newcommand{\Nr}{{\bf r}}
\newcommand{\NF}{{\bf F}}
\newcommand{\Nx}{{\bf x}}

\newcommand{\Ny}{{\bf y}}
\newcommand{\Nf}{{\bf f}}
\newcommand{\Nv}{{\bf v}}
\newcommand{\Nk}{{\bf k}}
\newcommand{\Nna}{{\bf \nabla}}
\newcommand{\NR}{{{\bf R}}}
\newcommand{\qq}{\begin{eqnarray}}

\newcommand{\da}{\partial}
\newcommand{\ee}{{\rm e}}

\newcommand{\qqq}{\end{eqnarray}}

\newcommand{\tr}{\hbox{tr}}

\newcommand{\CO}{{\cal O}}

\newcommand{\CT}{{\cal T}}

\newcommand{\s}{\hspace{0.05cm}}
\newcommand{\m}{\hspace{0.025cm}}

\newcommand{\dd}{{{}^-\hs{-0.3cm}d}}

\newcommand{\hf}{{_1\over^2}}
\newcommand{\ov}{\overline}

\begin{document}
\
\vskip 1.7cm
\begin{center}
{\large{\bf{EASY \ TURBULENCE}}}\m\footnote{lectures
given at the IX$^{\rm th}$ CRM Summer School ``Theoretical Physics
at the End of the XX$^{\rm th}$ Century'', Banff (Canada),
June 27 to July 10, 1999}
\vskip 1.4cm
Krzysztof Gaw\c{e}dzki
\vskip 0.3cm
C.N.R.S., I.H.E.S. 

91440 Bures-sur-Yvette, France
\end{center}

\date{ }

\addtocounter{section}{1}
\vskip 1.8cm


It seems a safe bet that the understanding of developed 
turbulence, a long standing challenge for theoretical 
and mathematical physics, will enter into the third
millennium as an unsolved problem. This is an introductory 
course to the subject. We discuss
\vskip 0.2cm

\no\hbox to 3.2cm{in \ {\bf{Lecture 1}}:\hfill}the Navier Stokes 
equations, existence of solutions, statistical description,
energy balance and cascade picture;
\vskip 0.2cm

\no\hbox to 3.2cm{in \ {\bf{Lecture 2}}:\hfill}the Kolmogorov 
theory of three-dimensional turbulence versus intermittency,
the Kraichnan-Batchelor theory of two-dimensional turbulence; 
\vskip 0.2cm

\no\hbox to 3.2cm{in \ {\bf{Lecture 3}}:\hfill}the Richardson 
dispersion law and the breakdown of the Lagrangian flow;
\vskip 0.2cm

\no\hbox to 3.2cm{in \ {\bf{Lecture 4}}:\hfill}direct and inverse 
cascades and intermittency in the Kraichnan model of passive 
advection. 
\vskip 2cm

\no {\large\bf{LECTURE 1}}
\vskip 0.5cm

Theoretical physics pursues two goals. On one side, it
searches for fundamental laws of nature. On the other side,
it studies the theoretical and phenomenological consequences 
of the laws already found. Hydrodynamics represents the second 
case. Its fundamental laws, in the form of the Navier-Stokes (NS)
equation or its variations, were formulated in the first
half of the nineteenth century by Navier (1823) and Stokes (1843)
as a modification of the even older Euler equation (1755).
The NS equation describes the temporal evolution of a {\bf velocity} 
field $\Nv(t,\Nx)$ in gasses or liquids. It takes the form
\qq
\partial_t\Nv+(\Nv\cdot\Nna)\Nv-\nu\Nna^2\Nv={_1\over^\rho}
(\Nf-\Nna p)\,,
\label{NS}
\qqq
where $\nu$ is the viscosity of the fluid ($\cong 1.5\times 10^{-5}\,
{\rm m^2\over sec}$ for air, $\cong 10^{-6}\,{\rm m^2\over sec}$
for water), $\rho$ is the fluid density, $\Nf$ is the external 
(intensive) force and $p$ is the pressure. In most physical 
applications, the dimensionality of the space is $3$ or $2$, but 
the equations make sense in a general dimension $d$. 
The Euler equation without the $\nu\Nna^2\Nv$ term 
is really the \s$\NF=m\Na\s$ (or rather $\Na={_1\over^m}\,\NF$\,) 
\,relation for the volume element of the fluid.
The $\nu\Nna^2\Nv$ term in the NS equation
represents the friction forces.
The relation (\ref{NS}) has to be supplemented 
with the continuity 
equation $\ \da_t\rho+\Nna\cdot(\rho\Nv)=0\,,$ \ and an equation 
of state relating $\rho$ and $p$. In most applications, one may 
assume that the fluid is {\bf incompressible}, i.e.\,\,that $\rho$ 
is constant and that $\Nv$ is divergence free\footnote{we shall 
absorb in this case the constant $1\over\rho$ into $p$ and $\Nf$}: 
\qq
\Nna\cdot\Nv=0\,. 
\label{div}
\qqq
It follows then by taking the divergence of both sides of 
Eq.\,\,(\ref{NS}) that $\Nna^2p=-\Nna\cdot(\Nv\cdot\Nna)\Nv$ 
so that pressure may be calculated for given velocity field.
It may be also eliminated by applying to the NS equation
the transverse projector which leaves 
the divergenceless $\Nv$ unchanged. 
The incompressible Euler equation has a nice infinite-dimensional 
geometric interpretation: it describes the geodesic flow on the
group of volume preserving diffeomorphisms\footnote{recall
that the Euler top is related to the geodesic flow on the group
$SO(3)$}. In one space dimension and without the incompressibility 
constraint and the pressure term, Eq.\,\,(\ref{NS}) becomes 
the Burgers equation, one of the simplest non-linear equations. 
\vskip 0.3cm

The incompressible Euler and NS equations are examples 
of nonlinear partial differential evolution equations. 
After a century and a half of studies, they still 
pose major open problems as far as the control of their 
solutions is concerned. The most interesting questions 
touch on the short-distance (ultra-violet) behavior. 
Suppose that we start from smooth initial data and 
the force $\Nf$ is smooth. For simplicity, let us 
assume compact support of both (we may also consider 
the compact space or the case with boundary conditions). It is 
known that the smooth solutions of the so posed initial
value problem are unique and exist for short time. 
Do they exist for all times? The answer is positive in 
2 dimensions for both Euler and NS equations but 
in 3 dimensions the answer is not known. It is usually
expected to be positive in the NS case. The opinions about
the Euler case (no blowup versus finite-time blowup 
for special smooth initial conditions) are more divided
and fluctuate in time.
\vskip 0.3cm

In an important 1933 paper on the NS equation, Leray has
introduced the notion of weak solutions of the equation. A 
vector field $\Nv(t,x)$ locally in $L^2$ is a weak solution if
it satisfies the equations in the distributional sense,
i.e.\,\,if for any smooth vector field $\Nu$ without divergence
and any smooth function $\varphi$, both with compact supports,
\qq
\int\left[\left(\da_tu^i+\nu\Delta u^i
+(\da_ju^i)v^j\right)v^i\s+\s u^if^i\right]\s=\s0\ \quad
{\rm and}\ \quad\int(\da_i\varphi)\m v^i\s=\s0\,.
\non
\qqq
Leray showed by compactness arguments existence of global weak 
solutions of the 3-dimensional NS equations with additional
properties (e.g. with space derivatives locally square
integrable). The weak solutions are not unique (there are weak 
solutions of the 2-dimensional Euler equation with compact 
support).
\vskip 0.5cm

The NS equation is invariant under rescalings. Let
\qq
&&\tilde\Nv(t,\Nx)\s=\s\tau\m s^{-1}\s \Nv(\tau t,s\Nx)\s,\cr
&&\tilde\Nf(t,\Nx)\s=\s\tau^2s^{-1}\s \Nf(\tau t,s\Nx)\s,\cr
&&\tilde p(t,\Nx)\s=\s\tau^2s^{-2}\s p(\tau t,s\Nx)\s,\cr
&&\tilde\nu\s=\s\tau\m s^{-2}\s\nu\s.
\non
\qqq
If $\Nv$ and $p$ solve the NS equation with viscosity $\nu$
and force $\Nf$ then $\tilde\Nv$ and $\tilde p$ give a solution
for viscosity $\tilde\nu$ and force $\tilde\Nf$.
It is convenient to introduce the dimensionless version 
of the (inverse) viscosity, the Reynolds number
\qq
Re\s=\s{L\ \delta_{_L}\hspace{-0.06cm}v\over\nu}\s,
\non
\qqq
where $\delta_{_L}\hspace{-0.06cm}v$ is a characteristic 
size of velocity differences 
over scale $L$ of the order of the size of the system. 
Note the scale-dependent nature of the concept.
For the flow in a pipe of radius $L$, we may 
take $\delta_{_L}\hspace{-0.06cm}v$ as velocity 
in the middle of the pipe (the velocity vanishes 
on the wall of the pipe). The following are the basic 
phenomenological observations about hydrodynamics. If $Re\ll 1$, 
one encounters regular ("{\bf laminar}") flows. For $Re$ 
between $\sim 1$ and $\sim 10^2$, one observes complicated 
phenomena depending on the precise situation. For $Re\gg 10^2$, 
very irregular ("{\bf turbulent}") flows set in. They show 
for high $Re$ ("{\bf developed turbulence}") a certain degree 
of similarity for different circumstances. 
\vskip 0.3cm

Somewhat simplifying, one could say
that for the laminar flows the non-linear term $(\Nv\cdot\Nna)\Nv$ 
of the NS equation plays a smaller role. This is
a relatively well understood regime, also rigorously.
Following Gallavotti's article cited at the end of the Lecture, 
define the running Reynolds number
\qq
Re_r\s=\s{r^2\over\nu}\m\left({1\over{\vert B_r\vert}}\int_{_{B_r}}
\hs{-0.3cm}\vert\Nna\Nv\vert^2\right)^{\hs{-0.1cm}1/2}
\non
\qqq
where \s$B_r\m=\m\{\m\s(s,\Ny)\ \vert
\ \vert s-t\vert<{r^2\over\nu}\s,
\ \vert \Ny-\Nx\vert< r\s\}\s$ is a neighborhood of the space-time
point $(t,\Nx)$. Note that we may rewrite
\qq
Re_r\s=\s{r\ \delta_rv\over \nu}
\non
\qqq
where $\delta_rv$, the mean velocity difference on scale 
$r$, is calculated by multiplying the mean square gradient 
of $v$ over $B_r$ by $r$. The best regularity result about 
the weak solutions of the NS equation is due 
to Caffarelli-Kohn-Nirenberg and says 
that there exists $\epsilon>0$ such that if $Re_r\leq\epsilon$ 
then the solution is smooth in the $B_{\epsilon\m r}$ 
neighborhood of $(t,\Nx)\m$. \s This implies that, for 
a weak solution, the Hausdorff dimension 
of the set of singularities is $\leq\m 1$. \m Note the 
spirit of the result in line with the phenomenological
characterization of laminar flows.
\vskip 0.3cm

One expects that in the regime of intermediate $Re$ 
between $\sim 1$ and $\sim 10^2$ only a finite number of modes 
of fluid play an important role. These modes 
may be effectively described
by ordinary differential equations to which the theory 
of dynamical systems (bifurcations, strange attractors, chaology)
may be applied. Indeed, the dynamical system 
theory has been used with much success to describe specific 
situations, as the flow between rotating cylinders, 
for example. It is not clear, however, if the 
dynamical system ideas may be useful to describe 
the fully developed turbulence. 
\vskip 0.3cm

The importance of the NS equations is far from 
being limited to the mathematical questions. It extends 
to meteorology, aeronautics and maritime engineering, to mention 
only three of the most important domains of practical 
applications. The regime of large $Re$, where the non-linear 
term of the NS equation becomes very important dominates 
in practical situations (in medium size river, $Re\sim 10^7$). 
One has to admit that the interests of engineers and 
theoretical physicists are somewhat different.
The first ones are interested mainly in flows
around obstacles (e.g.\,\,an airplane wing) whereas
the second ones show a tendency to concentrate on flows 
far from boundaries where the simplifying assumptions 
of homogeneity and isotropy seem in place. Nevertheless, 
a good understanding of such idealized flows would certainly 
have practical consequences.
\vskip 0.3cm

For high Reynolds numbers, it is reasonable to attempt
a statistical description of complicated turbulent flows.
In the theoretical approach, the statistics may be generated 
by considering random initial data or/and random forcing.
Since some degree of universality is observed in this situation 
independently on the way the flow is excited, one often assumes
that the force $\Nf$ is a random Gaussian field with mean zero 
and the covariance
\qq
\langle\m f^i(t,\Nx)\s f^j(s,\Ny)\rangle\s=\s\delta(t-s)\s\s 
\chi^{ij}({_{\Nx-\Ny}\over^L})
\label{rf}
\qqq
with $\da_i \chi^{ij}=0\s$ and the {\bf injection scale} 
$L$ regulating 
the decay of $\chi$, i.e. the scale on which the force acts.
$\Nv(t,\Nx)$ becomes then a random field and the NS equation 
takes, schematically, the form 
\qq
\da_t\Phi\s=\s-\m F(\Phi)\s+\s \eta\,.
\label{ee}
\qqq
\vskip 0.3cm

Many other dynamical problems in physics may be put in such 
a form with \s$F(\Phi)\s$ being a nonlinear functional
of local densities \s$\Phi(t,\Nx)\s$ of physical quantities and 
\s$\eta\s$ a random noise. The case of the 
NS equation should be contrasted with another example 
of Eq.\s\s(\ref{ee}), provided by the Langevin equation 
describing the approach to equilibrium in systems 
of statistical mechanics or field theory. 
In the latter case, the nonlinearity is of the gradient type:
\qq
F(\Phi)\s=\s{{\delta S(\Phi)}\over{\delta\Phi}}
\non
\qqq
with \s$S(\Phi)\s$ a local functional, 
e.g. $S(\Phi)=\hf\int(\nabla\Phi)^2+\hf m^2\int\Phi^2
+\lambda\int\Phi^4\s$ in the \s$\Phi^4\s$ field theory.
The noise is taken Gaussian: 
\qq
\langle\m \eta(t,\Nx)\s \eta(s,\Ny)\rangle\s=\s\delta(t-s)\s\s 
L^{-d}\,\chi({_{\Nx-\Ny}\over^L})\,.
\label{rf1}
\qqq
The covariance \s$L^{-d}\m\chi(\Nx/L)\s$ regulates the theory on short
distances \s$\mathop{<}\limits_{^\sim}\m L\s$ and is close to 
the delta-function \s$\delta(\Nx)\m$ for small \s$L\m$.
On the contrary, in the case of the NS equations we are mostly 
interested in forces acting on large distances $\sim L$
(e.g. the convective forces in the atmosphere) so that 
the force covariance \s$\chi^{ij}(\Nx/L)\s$ becomes close 
to a constant in the position space, i.e.\,\,to a multiple
of the delta-function 
in the wavenumber space. Such regime in field theory would 
correspond to distances shorter than the ultraviolet cutoff, 
with the behavior strongly dependent on the detailed form 
of the cutoff. Another difference is that in Eq.\s\s(\ref{NS}) 
the non-linear term, dominant for high Reynolds numbers, 
is not of the gradient type. Finally, the presence of pressure 
renders it also nonlocal, which is another complication. 
All these differences make the case of the NS equation quite 
different from that of the Langevin equation describing, 
in the stationary regime, equilibrium states. They make  
the NS problem, strongly coupled for high $Re$, resistant 
to the methods employed successfully in the study of equilibrium 
states like perturbative approaches or the renormalization group.
\vskip 0.3cm

The main characteristic of the stationary regimes of 
the randomly forced NS equations is the presence of
non-vanishing fluxes of conserved quantities, forbidden
in equilibrium states. By integrating the scalar product 
of the incompressible NS equation with $\Nv$ over the space 
and assuming that the flow velocity vanishes sufficiently 
fast at large distances (or at the boundary), one deduces 
the energy balance:
\qq
\da_t\s\m{_1\over^2}\int \Nv^2\s=\s-\m\nu\int(\Nna\Nv)^2
\s+\s\int\Nf\cdot\Nv\s. 
\label{bal}
\qqq
The equation states that the rate of change 
of fluid energy is equal to the energy injection rate
$\int\Nf\cdot\Nv$ (work of the external forces per unit time) 
minus the energy dissipation per unit time $\,\nu\hspace{-0.06cm}
\int(\Nna\Nv)^2$ due to the viscous friction. In a stationary state, 
the mean overall energy of the fluid is constant in time so that 
the energy balance equation (\ref{bal}) implies that
\qq
\int\langle\m\nu\m(\Nna\Nv)^2\rangle\s=\s
\int\langle\m\Nv\cdot\Nf\rangle\s,
\non
\qqq
where $\s\langle\ -\ \rangle$ denotes the ensemble
average, or, that in the homogeneous state, 
\qq
\bar\epsilon\s\equiv\s\langle\m\nu\m(\Nna\Nv)^2\rangle
\s=\s\langle\m\Nv\cdot\Nf\rangle\s\equiv\s\bar\varphi
\label{bal1}
\qqq
where $\bar\epsilon$ denotes the mean dissipation rate
and $\bar\varphi$ the mean injection rate of energy,
both with the dimension ${length^2\over time^3}$.
In the situation where the energy injection is a large
distance process (e.g. in the atmospheric turbulence or shear flows)
one expects that for high $Re$ a {\bf scale separation} occurs, 
with the energy dissipation taking place on much smaller
distances. Pictorially, energy is transmitted to
the fluid by the excitation of large eddies on scale $L$ 
which subsequently break to smaller scale eddies 
and so on. This way the injected energy is passed to shorter 
and shorter scales without substantial loss, until the viscous
scale $\eta$ is reached where the friction dissipates
energy. Such an {\bf energy cascade}, described first 
by Richardson in 1922, is characterized by the integral 
scale $L$, the viscous scale $\eta$ and the energy
dissipation rate $\bar\epsilon$. The scale ratio $L/\eta$ 
should grow with the Reynolds number. The interval of distance 
scales $r$ satisfying $L\gg r\gg\eta$ is
called the {\bf inertial range}. 
\vskip 0.3cm

The cascade picture may be formulated in more quantitative 
terms by introducing the quantities
\qq
\bar\epsilon_{\leq K}&=&\m\nu\int
\limits_{\vert\Nk\vert\leq K}\bigg(\int
\ee^{-i\s\Nk\cdot\Nx}\s\s\langle\m
\Nna\Nv(\Nx)\cdot\Nna\Nv({\bf{0}})\rangle
\,\,d\Nx\bigg)\s\dd\Nk\,,\cr
\bar\varphi_{\leq K}&=&\int\limits_{\vert 
\Nk\vert\leq K}\bigg(\int\ee^{-i\s \Nk\cdot\Nx}\,\,\langle\m
\Nv(\Nx)\cdot\Nf({\bf{0}})\rangle
\s\s d\Nx\bigg)\s\dd\Nk
\non
\qqq
interpreted as the mean dissipation and mean injection 
rate in wavenumbers $\m\Nk\m$ with $\m\vert\Nk\vert\leq K\,$
($\dd\Nk\equiv{d\Nk\over(2\pi)^{d}}$). 
The injection of energy limited to distances $\gsim L$
means that, as a function  of $K$, $\s\bar\varphi_{\leq K}\s$ 
is close to $\bar\epsilon$ everywhere except for $K\lsim{1\over L}$ 
where it falls to zero with $K\to0$. E.g., for the time-decorrelated
force with an appropriate interpretation of Eq.\,\,(\ref{NS}) as 
a stochastic differential equation, the relation 
$\m\langle\Nv(t,\Nx)\,\Nf(t,{\bf 0})\rangle
=\hf\,\tr\,\chi({\Nx\over L})\m$ holds so that 
$\,\bar\varphi_{\leq K}=\hf\int\limits_{\vert \Nk\vert\leq K}
(\int\ee^{-i\s\Nk\cdot\Nx}\,\,\tr\,\chi({_{\Nx}\over^L})
\s\s d\Nx\m)\s\dd\Nk\m$ and such behavior of 
$\s\bar\varphi_{\leq K}\s$ follows since $\m\chi\m$ is close
to a delta function in the wavenumber space. The cascade picture 
should imply that the mean dissipation rate 
$\bar\epsilon_{\leq K}$ is negligable
for $K\ll{1\over\eta}$ and then grows to $\bar\epsilon$.
The difference
\qq
\bar\varphi_{\leq K}-\bar\epsilon_{\leq K}\s\equiv\s
\bar\pi_{K}
\non
\qqq
has the interpretation of the energy flux out of the
wavenumbers $\Nk$ with $\vert\Nk\vert\leq K$. This flux should be
approximately constant and equal to $\bar\epsilon$ 
in the inertial range \s${1\over L}\ll K\ll{1\over\eta}\s$,
see Fig.\,\,1.
\vskip 0.6cm

\noindent{\bf Bibliography}
\vskip 0.3cm

\noindent For the original formulation of the NS equations see:
\vskip 0.1cm

1. \ C. L. M. H. Navier: {\it M\'{e}moire sur le lois 
du mouvement des fluides}.
M\'em. Acad. Roy. Sci. {\bf 6} (1823), 389-440
\vskip 0.1cm

2. \ G. G. Stokes: {\it On some cases of fluid motion}.
Trans. Camb. Phil. Soc. {\bf 8} (1843), 105
\vskip 0.3cm

\noindent For the Burgers equation see:
\vskip 0.1cm

3. \ J. M. Burgers: {\it The Nonlinear Diffusion Equation}.
D. Reidel, Dordrecht 1974
\vskip 0.3cm

\noindent A brief introduction to mathematical problems 
of the incompressible Navier-Stokes equations, from which we have 
borrowed the discussion of the regularity results, is
\vskip 0.1cm

4. \ G. Gallavotti: {\it Some rigorous results about 3D Navier-Stokes}.
Lecture Notes, Les Houches 1992, mp${}_{-}$arc/92105
\vskip 0.3cm

\noindent For a textbook on mathematical theory of NS equations, see:
\vskip 0.1cm

5. \ R. Temam: {\it Navier-Stokes Equations}. North-Holland, 
Amsterdam 1984
\vskip 0.3cm      

\noindent The dynamical system approach to turbulence is described 
e.g.\,\,in:
\vskip 0.1cm

6. \ D. Ruelle: {\it The turbulent fluid as a dynamical system}. 
In {\it New Perspectives in Turbulence}, ed. L Sirovich, Springer, 
Berlin 1991, pp. 123-138
\vskip 0.1cm

7. \ T. Bohr, M. H. Jensen, G. Paladin and A. Vulpiani: {\it
Dynamical System Approach to Turbulence}. Cambridge University Press,
Cambridge 1998
\vskip 0.3cm

\noindent The mathematical approach to the statistics of solutions 
is discussed in the textbook
\vskip 0.1cm

8. \ M. J. Vishik and A. V. Fursikov: {\it Mathematical Problems
of Statistical Hydrodynamics}. Kluwer, Dordrecht 1988
\vskip 0.3cm      

\noindent The classical textbooks on  fluid dynamics are:
\vskip 0.1cm

9. \ G. K. Batchelor: {\it Introduction to Fluid Dynamics}. Cambridge 
University Press, Cambridge 1967
\vskip 0.1cm

10. \,A. S. Monin and A. M. Yaglom: {\it Statistical Fluid Mechanics I \&
II}. MIT Press, Cambridge MA 1971 \& 1975
\vskip 0.3cm

\noindent For the original description of the turbulent cascade see:
\vskip 0.1cm

11. \,L. F. Richardson: {\it Weather Prediction by Numerical Process}.
Cambridge University Press, Cambridge 1922

\leavevmode\epsffile[-75 -20 345 200]{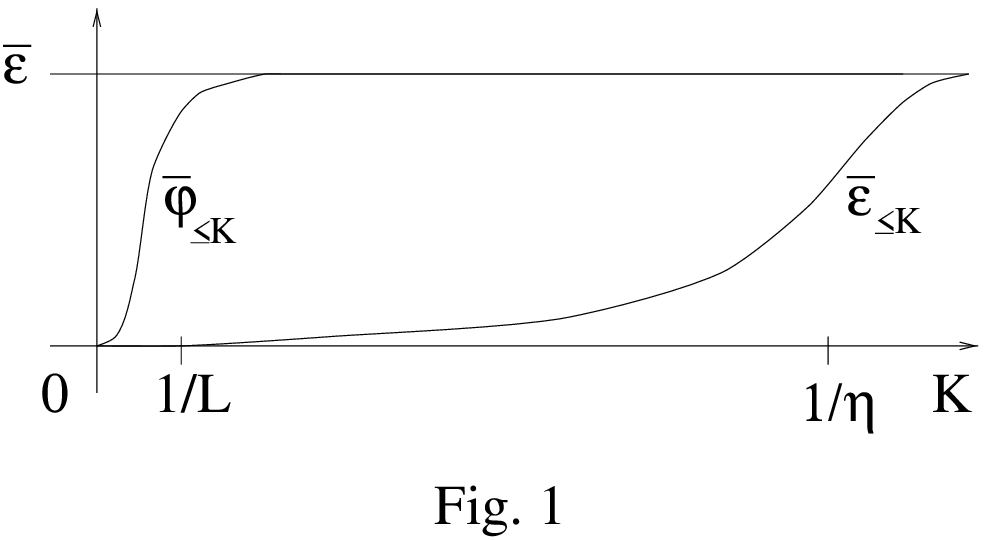}
\eject
\ \vskip 1cm

\no {\large\bf{LECTURE 2}}
\vskip 0.5cm

In 1941, A. N. Kolmogorov has proposed a scaling
theory of the developed turbulence which has deeply marked 
the ensuing turbulence research and rests a reference point 
of most of the modern research on the subject.
It is based on the exact statistical relations for the 
turbulent velocities obtained with general assumptions
and going back to the 1938 work of von K\'{a}rm\'{a}n and 
Horwarth. The NS equation implies that
\qq
\Nv(t+\Delta t)\s=\s\Nv(t)\s+\s[\m-\m(\Nv\cdot\Nna\Nv)\m\Nv
+\nu\Nna^2\Nv-\Nna p\m]\vert_{_{t}}\Delta t\s
+\s\int_{_t}^{^{t+\Delta t}}
\hs{-0.5cm}\Nf(s)\s\m ds\s+\s\CO((\Delta t)^2)
\non
\qqq
(the term involving the white noise in time 
$\Nf$ is $\CO(t^{1/2})$).
Consequently, we obtain for the statistical expectation of  
the scalar product of two velocities at the same time
but general points:
\qq
&&\langle\s\Nv(t+\Delta t,\Nx)\cdot\Nv(t+\Delta t,\Ny)
\s\rangle\s=\s\langle\s\Nv(t,\Nx)\cdot\Nv(t,\Ny)\s\rangle\cr\cr
&&+\Big[-\s\langle\s (\Nv\cdot\Nna)\Nv(\Nx)\cdot\Nv(\Ny)\s
\rangle\,-\s\langle\s\Nv(\Nx)\cdot 
(\Nv\cdot\Nna)\Nv(\Ny)\s\rangle \cr\cr
&&+\s\nu\langle\s\Nna^2\Nv(\Nx)\cdot\Nv(\Ny)\s\rangle
\s+\s\nu\langle\s\Nv(\Nx)\cdot\Nna^2\Nv(\Ny)\s
\rangle\s+\s\tr\s\chi({_{\Nx-\Ny}
\over^{L}})\Big]\Big|_{_t}\Delta t\s+\s\CO((\Delta t^2))\s.
\non
\qqq
We have dropped the pressure terms assuming the homogeneity
(i.e.\,\,translation invariance) of the statistical state.
The second line terms may be rewritten as
\qq
\m{_1\over^2}\s\Nna_\Nx\cdot\langle\m
(\Nv(\Nx)-\Nv(\Ny))\m(\Nv(\Nx)-\Nv(\Ny))^2\rangle
\non
\qqq
and the next two ones as
\qq
-2\nu\,\langle\m\Nna\Nv(\Nx)\cdot\Nna\Nv(\Ny)\m\rangle\,.
\non
\qqq
Equating the $\CO(\Delta t)$ terms, we obtain 
\qq
\da_t\,\langle\m\Nv(\Nx)\cdot\Nv(\Ny)\m\rangle&=&  
{_1\over^2}\s\Nna_\Nx\cdot\langle\m(\Nv(\Nx)-\Nv(\Ny))\m
(\Nv(\Nx)-\Nv(\Ny))^2\rangle\s\cr\cr
&&-\,2\m\nu\s\langle\m\Nna\Nv(\Nx)\cdot
\Nna\Nv(\Ny)\rangle
\,+\s\tr\s\s\chi({_{\Nx-\Ny}\over^L})\s,
\label{rel0}
\qqq
which is the basic relation between the 2-point and
the 3-point correlation functions of velocity.
\vskip 0.3cm

In three (and more) dimensions, we expect the stabilization
of the correlation functions for long times. In the stationary 
regime, the time derivatives of the equal-time functions 
vanish and we infer that
\qq
-{_1\over^4}\s\Nna_\Nx\cdot\langle\m(\Nv(\Nx)-\Nv(\Ny))\m
(\Nv(\Nx)-\Nv(\Ny))^2\rangle\,+\,\nu\s\langle\m\Nna\Nv(\Nx)\cdot
\Nna\Nv(\Ny)\rangle\ =\ {_1\over^2}\s
\tr\s\s\chi({_{\Nx-\Ny}\over^L})\s.
\label{rel}
\qqq
Taking first the limit $\Ny\to\Nx$ for positive $\nu$
and assuming that the presence of the latter smoothes out
the behavior of \s$\langle\m(\Nv(\Nx)-v(\Ny))(\Nv(\Nx)-\Nv(\Ny))^2
\m\rangle\s$ so that the first term on the left hand side vanishes
in the limit, we obtain
\qq
\bar\epsilon\s=\s{_1\over^2}\s\tr\s\s\chi({\bf 0})
\non
\qqq
which is nothing else but the energy balance equation
(\ref{bal1}) for the case of the time decorrelated force injecting 
energy at the mean rate $\bar\varphi=\hf\m\tr\,\chi({\bf 0})$. 
\vskip 0.3cm

To deduce further implications of Eq.\,\,(\ref{rel}), we take  
its invicid limit $\nu\to 0$ keeping the points separate. 
This gives
\qq
-\s{_1\over^4}\s\Nna_\Nx\cdot\langle\m(\Nv(\Nx)-\Nv(\Ny))\m
(\Nv(\Nx)-\Nv(\Ny))^2\rangle\vert_{_{\nu=0}}\s=\s{_1\over^2}\s\tr\s\s
\chi({_{\Nx-\Ny}\over^L})\s.
\label{rel1}
\qqq
The assumption that the force acts only on distances $\gsim L$
means that, for $\vert\Nx-\Ny\vert\ll L\m$, \s$\chi({_{\Nx-\Ny}\over^L})
=\chi({\bf 0})$, \m approximately, so that Eq.\s\s(\ref{rel1}) 
implies that in the inertial range,
\qq
-\s{_1\over^4}\s\Nna_{_{\Nx}}\cdot\langle\m(\Nv(\Nx)-\Nv(\Ny))^2
(\Nv(\Nx)-\Nv(\Ny))\m\rangle\s=\s\bar\epsilon\s.
\non
\qqq
Assuming the isotropy (i.e.\,\,the rotation invariance), this implies
the relation 
\qq
\langle\m(v^i(\Nx)-v^i(\Ny))\m(v^j(\Nx)-v^j(\Ny))\m
(v^k(\Nx)-v^k(\Ny))\m\rangle
=-\m{_{4\s\bar\epsilon}\over^{d(d+2)}}
\s(\delta^{ij} r^k+\delta^{ik} r^j+\delta^{jk} r^i)\,,
\label{3pt}
\non
\qqq                                                        
where $\Nr=\Nx-\Ny$. In other words, the 3-point function 
of equal-time velocity difference is in the inertial range
linear in the point separation. 
In particular, for the so called longitudinal 3-point structure function,
we obtain
\qq
S^\Vert_3(r)\ \equiv\ \langle\m(\Nv(\Nx)-\Nv(\Ny))\cdot{_\Nr\over^r})^3
\m\rangle\ =\ -\m{_{12}\over^{d(d+2)}}\,\bar\epsilon\, r
\label{S3}
\qqq
known, for $d=3$, as the {\bf Kolmogorov four-fifths law}.
\vskip0.3cm

One may deduce a stronger version of the above relation which 
takes the form of the operator product expansion
for the \s$\nu\to0\s$ limit of the dissipation operator 
\s$\epsilon=\nu\m(\nabla\Nv)^2\m$:
\qq
\epsilon(\Nx)\ =\ -\m{_1\over^4}\s\m\lim\limits_{\Ny\to\Nx}\s
\s\Nna_{_\Nx}\cdot[(\Nv(\Nx)-\Nv(\Ny))\m
(\Nv(\Nx)-\Nv(\Ny))^2]\s\bigg\vert_{{\nu=0}}
\label{da}
\qqq
which should hold inside expectations in the \s$\nu\to0\s$ limit. 
As noticed recently by Duchon and Robert, the relation (\ref{da})
holds for all weak solutions of the Euler equation which are limits 
of strong solutions of the NS equation. Relation (\ref{da}) 
is often called a {\bf dissipative anomaly}:
the dissipation \s$\epsilon\s$ whose definition involves
a factor of \s$\nu\s$ does not vanish when \s$\nu\to0\m$.
\vskip 0.3cm

In his 1941 paper, Kolmogorov went one step further 
by postulating the universal character of the turbulent
cascade in the inertial range with the equal-time correlators 
of velocity differences over inertial range distances given 
by universal functions of the latter and of the dissipation rate 
$\bar\epsilon$. In particular this implies that the velocity 
structure functions $\,S^\Vert_n(r)=\langle(\Nv(\Nx)-\Nv(\Ny))
\cdot{\Nr\over r})^n\rangle\,$ are determined 
by dimensional reasons up to universal constants:
\qq
S^\Vert_n(r)\s=\s C_n\s\bar\epsilon^{n/3}\m r^{n/3}\,.
\label{K41}
\qqq
Indeed, the right hand side is the only function 
of $\bar\epsilon$ and $r$ with the dimension $({length\over 
time})^n$. \,The physical content of the Kolmogorov theory is 
that the typical velocity $v_r$ of size $r$ eddies behaves as 
$\,{\bar\epsilon}^{1/3}\m r^{1/3}\m$. \,During the {\bf eddy 
turnover time} $\,\tau_r={r\over v_r}\propto\bar\epsilon^{-1/3}
\m r^{2/3}\m$, \,these eddies transfer their energy with 
the density $\,\hf\m v_r^2\propto{\bar\epsilon}^{2/3}\m r^{2/3}\,$ 
to the shorter scale resulting in the constant energy flux 
$\,\propto\bar\epsilon$. For the scale-dependent Reynolds 
number one obtains then \s$Re_r\propto{{\bar\epsilon}^{1/3}
\m r^{4/3}\over\nu}\m$. \s In particular, $\,Re=Re_L\propto
{{\bar\epsilon}^{1/3}\m L^{4/3}\over\nu}\s$ and \s$1=Re_\eta
\propto{{\bar\epsilon}^{1/3}\m \eta^{4/3}\over\nu}$. \m Hence 
\s${\eta\over L}\propto R^{-3/4}\s$ and it decreases with $Re$.
\vskip 0.4cm

As we have seen, the $n=3$ relation (\ref{K41}) 
coincides with the Kolmogorov four-fifths law. 
The structure functions are measured, more or less 
directly, in atmospheric or ocean flows, in water jets,
in aerodynamic tunnels or in subtle experiments
with helium gas in between rotating cylinders or plates.
They are also accessible in numerical simulations.
One extracts then the scaling exponents assuming
the behavior
\qq
S^\Vert_n(x)\s\propto\s r^{\zeta_n}\s.
\non
\qqq
$\zeta_3$ agrees well with the theoretical prediction
$\zeta_3=1$. Here are some other exponents obtained
from wind tunnel data 
\qq
\zeta_2=.70\s(.67)\m,\quad\hs{-0.06cm}\zeta_4=1.28\s(1.33)\m,
\quad\hs{-0.06cm}\zeta_5=1.53\s(1.67)\m,\quad\hs{-0.06cm}
\zeta_6=1.77\s(2)\m,\quad\hs{-0.06cm}\zeta_7=2.01\s(2.33)
\non
\qqq
with the Kolmogorov values in the parenthesis for comparison.
The discrepancy is quite pronounced (its direction for
the even functions is determined by the H\"{o}lder inequality
implying that $\zeta_n$ is a concave function of $n$). 
One of the main open problems
in the theory of fully developed turbulence is to explain,
starting from the first principles (i.e.\s\s from the NS equation),
the breakdown of the Kolmogorov theory leading to the anomalous 
structure-function exponents which indicate that the distribution
of $\Nv(t,\Nx)$ in the inertial range is rather different 
from Gaussian. The discrepancy may be measured by the {\bf skewness} 
$S_3/S_2^{3\over 2}$ or the {\bf flatness} $S_4/S_2^2$ 
which grow with diminishing distance instead of being equal
to their Gaussian values 0 and 3. Hence the domination 
of the short scales by the large deviations
of the velocity differences indicating an enhanced 
short-distance activity: the phenomenon called 
{\bf intermittency}. 
The intuitive explanation of intermittency which was 
advanced is that only a part of the fluid modes (temporal 
or/and spatial) participates in the turbulent cascade, 
with the proportion of active modes decreasing with 
diminishing scale. This forces the active short-distance 
modes to transfer more energy and, consequently, 
to be more excited. This picture led to multiple (multi)fractal
models of the cascade reviewed in the book by Frisch 
cited at the end of this Lecture. Such models, although 
interesting phenomenologically, are not based on the NS 
equation and allow to obtain essentially arbitrary spectra 
of exponents. They do not really explain the mechanism 
of the breakdown of the normal scaling in realistic flows.
\vskip 0.4cm

For the {\bf energy spectrum} 
\qq
\bar{e}_K\s\equiv\s{_1\over^2}\s{_d\over^{dK}}
\int\limits_{\vert \Nk\vert\leq K}\hs{-0.2cm}
\bigg(\int\langle\m\Nv(\Nx)\cdot\Nv(0)\rangle\s\s\ee^{-i\s\Nk\cdot\Nx}
\s\s d\Nx\bigg)\s\dd\Nk\s,
\non
\qqq
the Kolmogorov theory predicts
\qq
\bar{e}_K\propto\s{\bar\epsilon}^{2/3}\s K^{-5/3}
\label{5/3}
\qqq
for \s${1\over L}\ll K\ll{1\over\eta}\m$ just by the dimesional 
count. The experimental data seem to confirm this behavior (with
the possibility of a slight discrepancy consistent with the value 
of $\zeta_2$ cited above). For the mean dissipation rate, 
we obtain: $\,\bar{\epsilon}_{\leq K}=2\nu\int\limits_0^K
{K'}^2\m\bar{e}_{K'}\m dK'\propto\nu\,{\bar{\epsilon}}^{2/3}\m 
K^{4/3}\m$ confirming that $\m\bar{\epsilon}_{\leq K}/\bar{\epsilon}
=Re_{_{K^{-1}}}^{\,-1}\ll Re_\eta^{\m-1}=1\m$ in the inertial range. 
Deep in the dissipative regime $K\gg{1\over\eta}$, \s$\bar{e}_K$ 
falls off much
faster than in the inertial interval.
\vskip 1cm

An important object in the turbulence theory is the {\bf vorticity}
field that measures the strength and the orientation of eddies. 
In three dimensions it is a (pseudo-)vector field 
$\Nom=\Nna\times\Nv$ and it satisfies, for the incompressible 
flow, the equation 
\qq
\da_t\Nom+(\Nv\cdot\Nna)\Nom-(\Nom\cdot\Nna)\Nv
-\nu\Nna^2\Nom=\Nna\times\Nf\,.
\label{3vort}
\qqq
Note that $\,(\Nv\cdot\Nna)\Nom-(\Nom\cdot\Nna)\Nv\,$ is
the commutator of two vector fields, so that, at vanishing
$\nu$ and $\Nf$, the equation (\ref{3vort}) implies that
the vorticity is transported by velocity field as a vector.
This involves stretching of $\m\omega$ by the velocity 
{\bf strain} $\hf(\da_iv^j+\da_j v^i)$, an important mechanism 
for the energy transfer between scales. In two dimensions, 
however, the vorticity reduces to a (pseudo-)scalar field
$\omega=\epsilon^{ij}\da_i v^j$ whose evolution is
governed by the equation
\qq
\da_t\omega+(\Nv\cdot\Nna)\omega-\nu\Nna^2\omega=\epsilon^{ij}
\da_i f^j\,.
\label{3vort2}
\qqq
Here, at vanishing $\nu$ and $\Nf$, the vorticity is simply transported 
by the velocity field along the {\bf Lagrangian trajectories}
$\Nx(t)$ of the (imaginary) fluid particles s.t. ${d\Nx\over dt}=\Nv(t,\Nx)$.
\,In particular, the two-dimensional flow, unlike the tree-dimensional 
one, conserves, besides energy, the {\bf enstrophy} 
$\Psi=\hf\int\omega^2$ (as well as the integrals of higher 
powers of vorticity).
\vskip 0.3cm

In the seminal 1968 paper, R. H. Kraichnan has realized that
the conservation of enstrophy implies a very different
cascade picture in two dimensions, as compared to the
three-dimensional one, see also the paper of Batchelor of 1969. 
First of all, the {\bf enstrophy spectrum} is related 
to the energy spectrum:
\qq
\bar{\psi}_K\s\equiv\s{_1\over^2}\s{_d\over^{dK}}
\int\limits_{\vert \Nk\vert\leq K}\hs{-0.2cm}
\bigg(\int\langle\m \omega(\Nx)\,\omega({\bf{0}})
\rangle\s\s\ee^{-i\s \Nk\cdot\Nx}
\s\s d\Nx\bigg)\s\dd\Nk\s=\,K^2\,\bar{e}_K\,.
\non
\qqq
Similary, the mean enstrophy flux out of wavenumbers $\Nk$ with 
$\vert\Nk\vert\leq K$, if local in the momentum space, 
would have to be equal $K^2$ times the energy flux $\bar\pi_K$
so that the constancy of both fluxes is impossible. Kraichnan
reasoned that, in such a situation, it will be the enstrophy
flux which is constant on scales smaller than the injection  
scale so that the small scales will support a {\bf direct 
enstrophy cascade} towards smaller and smaller distances, 
with all the enstrophy dissipation taking place on the shortest 
scales. The energy flux towards small scales will then be damped 
and, as a result, the energy will be, instead, transferred to 
scales longer than the injection scale $L$ in an {\bf inverse 
energy cascade} process. If not impaired by boundaries or large 
scale friction, this process would lead to the pumping of energy 
into the constant mode at the rate equal to $\hf\m\tr\,\chi
({\bf{0}})$. 
\vskip 0.3cm

If we assume that $\ \langle\m\Nv(t,\Nx)\cdot\Nv(t,\Ny)\rangle-
t\,\,\tr\,\chi({\bf{0}})\ $ stabilizes at long times, 
then Eq.\,\,(\ref{rel0}) will lead to the relation
\qq
{_1\over^4}\s\Nna_\Nx\cdot\langle\m(\Nv(\Nx)-\Nv(\Ny))\m
(\Nv(\Nx)-\Nv(\Ny))^2\rangle\s-\,\nu\s\langle\m\Nna\Nv(\Nx)\cdot
\Nna\Nv(\Ny)\rangle\,=\s{_1\over^2}\s
\tr\s\s[\m\chi({\bf{0}})-\chi({_{\Nx-\Ny}\over^L})\m]\,.
\label{rel2}
\qqq
In particular, in the limit $\nu\to0$ and for $\vert\Nx-\Ny\vert
\gg L$,
\qq
{_1\over^4}\s\Nna_\Nx\cdot\langle\m(\Nv(\Nx)-\Nv(\Ny))\m
(\Nv(\Nx)-\Nv(\Ny))^2\rangle\s\,=\s{_1\over^2}\s
\tr\s\s\m\chi({\bf{0}})
\non
\qqq
implying the relation similar to the two-dimensional
version of Eq.\,\,(\ref{3pt}) but with the inverted sign.
In particular, we obtain
\qq
S^\Vert_3(r)\ =\ \m{_3\over^2}\,\bar\epsilon\, r\quad\quad{\rm for}
\quad\ r\gg L\,,
\non
\qqq
a {\bf three-halfs law}. 
\vskip 0.3cm

The inverse cascade of the two-dimensional turbulence has
been recently under an intensive study.
Both experimental and numerical data confirm the above prediction
and indicate that in this regime all structure functions, 
although non-Gaussian, scale with the Kolmogorov exponents 
$\zeta_n={n\over 3}$ indicating that the {\bf inverse cascade
is not intermittent}. This implies for the energy spectrum, 
the behavior (\ref{5/3}) for $K\ll{1\over L}$.
\vskip 0.3cm

For $\vert\Nx-\Ny\vert\ll L$, \,Eq.\,\,(\ref{rel2}) reduces
in the $\nu\to0$ limit to the relation
\qq
{_1\over^4}\s\Nna_\Nx\cdot\langle\m(\Nv(\Nx)-\Nv(\Ny))\m
(\Nv(\Nx)-\Nv(\Ny))^2\rangle\s=\s{_1\over^2}\s
\tr\s\s[\m\chi({\bf{0}})-\chi({_{\Nx-\Ny}\over^L})\m]\ 
=\ \CO(r^2)
\label{3ptd}
\qqq
implying the scaling  
\qq
S^\Vert_3(r)\ \sim\ r^3\quad\quad{\rm for}\quad\ r\ll L\,,
\non
\qqq
i.e.\,\,in the direct cascade regime. It is possible to infer
in two dimensions another exact relation for the 3-point 
functions by proceeding from Eq.\,\,(\ref{3vort2}) the same way 
that we did before. One obtains
\qq
\hf\,\da_t\,\langle\m\omega(t,\Nx)\,\omega(t,\Ny)\rangle&=& 
{_1\over^4}\,\Nna_{_\Nx}\cdot\langle\m(\Nv(t,\Nx)-\Nv(t,\Ny)\,
(\omega(t,\Nx)-\omega(t,\Ny))^2\rangle\cr\cr
&&-\,\nu\,\langle\m\Nna\omega(t,\Nx)
\cdot\Nna\omega(t,\Ny)\rangle\,-\,\hf\m\Nna_{\Nx}^2\,\tr\,\,
\chi({_{\Nx-\Ny}\over^L})\,.
\label{relo}
\qqq
Our previous assumption about the stabilization of the velocity
2-point function modulo a growing constant implies that
the 2-point function of the vorticity reaches a stationary
regime. We deduce from this, as before for the three-dimensional 
velocities, that the mean dissipation rate of vorticity
\qq
\bar\epsilon_\omega\,\equiv\,\nu\,\langle\m(\Nna\omega)^2\rangle
\,=\,
-\,\hf\m\Nna_{\Nx}^2\,\tr\,\,
\chi({\bf{0}})
\non
\qqq
is $\nu$-independent and that, in the limit $\,\nu\to0$,
\qq
-\m{_1\over^4}\,\Nna_{_\Nx}\cdot\langle\m(\Nv(t,\Nx)
-\Nv(t,\Ny)\,
(\omega(t,\Nx)-\omega(t,\Ny))^2\rangle\,
\,=\,-\,\hf\m\Nna_{\Nx}^2\,\tr\,\,
\chi({_{\Nx-\Ny}\over^L})\,.
\label{str}
\qqq
As noticed in a recent paper of Bernard, see the bibliography,
the relations (\ref{3ptd}) and (\ref{str}), both consistent 
with the $\CO(r^3)$ behavior of the 3-point functions
of velocity differences in the direct cascade
regime, allow to find exactly its asymptotic form.
If we assume next that the $n$-point functions of the velocity 
differences scale accordingly, i.e.\,\,with 
the power $n$, then, for $n=2$, we infer the Kraichnan-Batchelor
energy spectrum in the direct cascade
\qq
\bar{e}_K\ \sim\ K^{-3}\quad\quad{\rm for}\quad\ K\gg{_1\over^L}\,,
\label{KB}
\qqq
confirmed by experimental observations.
Intermittency, if existent in two-dimensional turbulence, seems 
much smaller then in three dimensions, especially in the
inverse cascade. There are, however, theoretical predictions 
of logarithmic corrections to the power-law scaling in the direct 
cascade, not yet accessible to experimental or numerical verification.
\vskip 0.5cm

{\bf Summarizing}: In three dimensions, in the inertial range 
of the direct (short distance) energy cascade, the 3-point velocity 
structure function scales linearly in the distance. The energy 
spectrum is close to $\propto K^{-5/3}$, with the anomalous 
scaling of higher-point structure functions signaling intermittency. 
In two dimensions, in the direct (short distance) enstrophy cascade, 
the 3-point structure function scales as the $3^{\rm rd}$ power
of the distance and the energy spectrum is close to $\,\propto K^{-3}$.
In the (long distance) inverse energy cascade, one observes 
the Kolmogorov scaling, similarly as in the three dimensional 
direct cascade, but with reduced intermittency. Assuming the above 
spectra, one infers that both in two and in three dimensions, 
the mean enstrophy density given by $\int K^2\m \bar{e}_K\,dK$ 
diverges in the ultraviolet in the invicid limit $\nu\to0$. 
Accordingly, one should expect that the stationary fully turbulent 
state is carried by weak solutions of the Euler equation 
with divergent enstrophy. Hence the importance of studying such 
solutions which, in three dimensions, dissipate energy on short 
distances by a mechanisms which is the topic of next Lecture.
\vskip 0.6cm

\noindent{\bf Bibliography}
\vskip 0.3cm

\noindent The first exact statistical relations
in turbulent flows were obtained in:  
\vskip 0.1cm

1. \ T. von K\'{a}rm\'{a}n and L. Horwarth: {\it On the statistical
theory of isotropic turbulence}. Proc. R. Soc. London {\bf A 164}
(1938), 192-215
\vskip 0.3cm

\noindent Kolmogorov's original work is  
\vskip 0.1cm

2. \ A. N. Kolmogorov: {\it The local structure of turbulence
in incompressible viscous fluid for very large Reynolds'
numbers}. C. R. Acad. Sci. URSS {\bf 30} (1941), 301-305
\vskip 0.3cm

\noindent A nice textbook reviewing the Kolmogorov theory 
of turbulence and much more, from which we have borrowed  
the discussion of the energy cascade and of the four-fifths 
law, is:
\vskip 0.1cm

3. \ U. Frisch: {\it Turbulence: the Legacy of A. N. Kolmogorov}.
Cambridge University Press, Cambridge 1995
\vskip 0.3cm      

\noindent The dissipative anomaly in weak solutions of the 
NS and Euler equations has been discussed in:
\vskip 0.1cm

4. \ J. Duchon and R. Robert: {\it Inertial energy dissipation 
for weak solutions of incompressible Euler and Navier-Stokes
equations}. 1999 preprint, submitted to Nonlinearity
\vskip 0.3cm      

\noindent For the data about the anomalous scaling 
in three-dimensional turbulence see:
\vskip 0.1cm

5. \ R. Benzi, S. Ciliberto, C Baudet and G. Ruiz Chavaria: 
{\it On the Scaling of Three Dimensional Homogeneous
and Isotropic Turbulence}. Physica {\bf D 80} (1995), 385-398
\vskip 0.3cm

\noindent The original works on the two-dimensional turbulence are:
\vskip 0.1cm

6. \ R. H. Kraichnan: {\it Inertial ranges in two-dimensional
turbulence}. Phys. Fluids {\bf 10} (1967), 1417-1423
\vskip 0.1cm

7. \ G. K. Batchelor: {\it Computation of the energy spectrum
in homogeneous two-dimensional turbulence}. Phys. Fluids
Suppl. II {\bf 12} (1969), 233-239
\vskip 0.3cm      

\noindent Our discussion of the 3-point functions
in two-dimensional turbulence follows:
\vskip 0.1cm

8. \ D. Bernard: {\it On the three point velocity correlation 
functions in 2d forced turbulence}. chao-dyn/9902010
\vskip 0.3cm

\noindent For recent experimental results on, respectively, 
the inverse and the direct cascades, see:
\vskip 0.1cm

9. \ J. Paret and P. Tabeling: {\it Intermittency in the 2D inverse
cascade of energy: experimental observations}. Phys. Fluids {\bf 10}
(1998), 3126-3136
\vskip 0.1cm

10. \,J. Paret, M. C. Jullien and P. Tabeling: {\it Vorticity 
statistics in the two-dimensional enstrophy cascade}. 
cond-mat/9904044
\vskip 0.3cm

\noindent For a recent numerical work on the inverse cascade
see:
\vskip 0.1cm

11. \,G. Boffetta, A. Celani and M. Vergassola: {\it Inverse cascade
in two-dimensional turbulence: deviations from Gaussianity}.
chao-dyn/9906016
\eject
\ \vskip 1cm

\no {\large\bf{LECTURE 3}}
\vskip 0.5cm

One of the first quantitative laws of fully developed
turbulence has been formulated by L. F. Richardson in 1926.
Basing on observations, e.g.\,\,of the movements of balloon probes
released in the atmosphere, Richardson noted that the
mean rate of growth of the square of the separation between 
two such probes is proportional to the four-thirds power
of the distance, instead of being distance-independent, as 
in the Brownian diffusion. In other words, the Richardson 
dispersion law states that
\qq
{d\rho^2\over dt}\ \propto\ \rho^{4/3}\,,
\label{RL}
\qqq
where $\rho(t)$ is the time $t$ distance between two
Lagrangian trajectories satisfying the ordinary differential 
equation (ODE)
\qq
{d\Nx\over dt}\ =\ \Nv(t,\Nx)\,.
\label{ODE}
\qqq
Of course, the Richardson law is compatible with the Kolmogorov
theory that implies that the velocity differences $\,\vert\Nv(t,\Nx)
-\Nv(t,\Ny)\vert\,$ scale as $\m\rho^{1/3}\m$ for $\m\rho=\vert\Nx
-\Ny\vert\m$ in the inertial range, but it deserves a more detailed 
discussion.
\vskip 0.3cm

Let us note that Eq.\,\,(\ref{RL}) is solved by 
\qq
\rho^{2/3}\ =\ \rho_0^{2/3}\,+\,C\, t
\non
\qqq
which, in the limit when the initial (i.e.\,\,time zero)
separation goes to zero, reduces to the relation 
\qq
\rho^2\ \,\propto\,\ t^3\,.
\label{prt}
\qqq
Of course, some care is needed since the law (\ref{RL}) 
was observed for separations in the inertial range, 
so for $\rho\gg\eta$, where $\eta$ is the viscous scale.
In the limit $Re\to\infty$, however, $\eta$ tends to zero 
and the growth (\ref{prt}) should hold for arbitrarily short 
separations. In the turbulent atmosphere, for example,
the viscous scale $\eta$ is of the order of millimeter so 
that neglecting it when we look at scales of meters
or even kilometers is a reasonable approximation. We infer  
this way that at $Re=\infty$, 
\vskip 0.2cm

{\bf infinitesimally close Lagrangian trajectories
separate in finite time}.
\vskip 0.4cm

Of course, $\rho$ is the average separation, with 
the mean taken over different velocity fields of the statistical
turbulent ensemble. One expects, however, that the trajectory 
separation is a self-averaging quantity so that this type 
of behavior holds already in a fixed typical velocity 
realization if we average over initial positions and times 
of release of two very close trajectories.
What is this strange behavior of the trajectories? 
\,We are not used to such behaviors of the solutions 
of the ODE's. For example, in the integrable dynamical 
systems the distance between two trajectories
\qq
\rho\ \simeq\ \CO(1)\,.
\non
\qqq
In the dissipative systems,
\qq
\rho\ \simeq\ \CO(\ee^{-\lambda t})\,.
\non
\qqq
In the chaotic dynamical systems,
\qq
\rho\ \simeq\ \CO(\ee^{\lambda t})\,,
\label{LE}
\qqq
with the Lyapunov exponent $\lambda>0$. Even the last case,
where the nearby trajectories separate fast, is quite different 
from the behavior (\ref{prt}). Indeed, for the exponential 
separation (\ref{LE}), infinitesimally 
close trajectories keep shadowing each other and never 
separate to a finite distance. Definitely, chaos and
fully developed turbulence are very different phenomena.
\vskip 0.3cm

In fact the {\bf explosive separation} (\ref{prt}) has a quite 
dramatic consequence: it means that, when $Re\to\infty$, 
the very concept of a Lagrangian trajectory determined
by its initial position in a fixed velocity realization breaks 
down. This breakdown of the Lagrangian flow is related 
to a breakdown of the theorem about 
the existence and unicity of solutions of the ODE
(\ref{ODE}). The theorem requires $\Nv(t,\Nx)$ to be Lipschitz
in $\Nx$, i.e.\,\,$\vert\Nv(t,\Nx)-\Nv(t,\Ny)\vert\sim\vert\Nx
-\Ny\vert\,$, \,whereas at $Re=\infty$ the velocities are only
H\"{o}lder continuous: \,$\vert\Nv(t,\Nx)-\Nv(t,\Ny)\vert\sim
\vert\Nx-\Ny\vert^{\alpha}\,$ with the exponent $\alpha<1$
($\alpha\simeq{1\over 3}$). We expect that for 
such velocities one may still maintain a probabilistic 
description of Lagrangian trajectories with such objects 
as the probability distribution function (PDF) 
$\,P^{t,s}(\Nx,\Ny\vert\Nv)\,$ of the time $s$ position 
$\Ny$ of the trajectory starting at time $t$ at point $\Nx$ 
still making sense. $\,P^{t,s}(\Nx,\Ny\vert\Nv)\,$ 
will rather be diffuse, however, than concentrated at one $\Ny$. 
In other words, we expect that, at $Re=\infty$, 
the Lagrangian trajectories become stochastic already 
in a fixed realization of the velocity field, providing 
a mechanism for the energy dissipation in weak solutions
of the Euler equation. 
\vskip 0.3cm

This important idea which seems to be a direct 
consequence in the limit of high Reynolds numbers  
of the Richardson dispersion law or of the Kolmogorov 
scaling of velocity differences has been 
rarely stressed in the past. It has appeared in a study
of weak solutions of the Euler equations (in a somewhat
different form) and in a recent analytic study 
of the Lagrangian trajectories in a simple statistical 
ensemble of velocities with the spatial H\"{o}lder continuity 
of the typical realizations built in. We shall spend the rest 
of this Lecture by reviewing the latter results with the aim 
to substantiate the preceding discussion. 
\vskip 0.4cm

Following R. H. Kraichnan who initiated in 1968 the study
of transport properties of velocities decorrelated in time,
see next Lecture, we shall consider a Gaussian ensemble
of velocities with mean zero and 2-point function
given by
\qq
\langle\s v^\alpha(t,\Nx)\s v^\beta(s,\Ny)\s\rangle\ =\ 
\delta(t-s)\s\s(D_0\delta^{\alpha\beta}-\m d^{\alpha\beta}
(\Nx-\Ny)\m)
\label{v2pt}
\qqq
with \s$D_0\s$ a constant and \s$d^{\alpha\beta}(\Nx)\m
\propto\m r^\xi\s$ for small $\m r\equiv\vert\Nx\vert\m$.
\,$0<\xi<2$ will be the parameter of the ensemble. The constant 
$D_0$ drops out in the correlations of the velocity differences. 
E.g.
\qq
\langle\,(v^\alpha(t,\Nx)-v^\alpha(t,{\bf 0}))\,
(v^\beta(s,\Nx)-v^\beta(s,{\bf 0}))\m\rangle\ =\ 
\delta(t-s)\,(d^{\alpha\beta}(\Nx)+d^{\beta\alpha}(\Nx))\,.
\label{fore}
\qqq
One may take 
\qq
D_0-d^{\alpha\beta}(\Nx)\ =\ \int{{\ee^{\m i\m\Nk\cdot\Nx}}\over
{(\Nk^2+L^{-2})^{(d+\xi)/2}}}\,\,\left(A\m{_{k^\alpha k^\beta}
\over^{\Nk^2}}\m+\m B\m(\delta^{\alpha\beta}-{_{k^\alpha k^\beta}
\over^{\Nk^2}})\right)\,\,\dd\Nk
\label{msr}
\qqq
with the infrared cutoff $L$. For $A=0$, the typical velocities 
are incompressible: $\Nna\cdot\Nv=0$, whereas for $B=0$
one obtains potential flows: $\Nv=\Nna\phi$. It is convenient 
to characterize the resulting velocity ensemble by, besides $\xi$, 
the {\bf compressibility degree}
\qq
\wp\ =\ {{\langle(\Nna\cdot\Nv)^2\rangle}
\over{\langle(\Nna\Nv)^2\rangle}}\ =\ {1\over{1+(d-1){B\over A}}}
\non
\qqq
contained between $0$ and $1$. \m$\wp=0$ corresponds to the
incompressible case whereas $\wp=1$ to the potential one.
\vskip 0.3cm

The above ensemble is not very realistic in its assumption
of temporal velocity decorrelation. Recall, however, that the eddy
turnover time $\tau_r$ is predicted by the Kolmogorov
theory to behave as $\,\sim r^{2/3}$ so that the time correlation
of the short scale eddies in small. The Kraichnan ensemble builds 
in the scaling behavior of the velocities in space
with the H\"{o}lder continuity $\,\vert\Nv(t,\Nx)-\Nv(t,\Ny)
\vert\sim\vert\Nx-\Ny\vert^{\xi/2}\,$ of typical realizations,
up to logarithmic corrections. To compare to the realistic 
ensemble of turbulent velocities, note that, for small $r$,
the right hand side of Eq.\,\,(\ref{fore}) behaves as 
\qq
{\rm const}.\,\,\delta({_{t-s}\over^{r^{2-\xi}}})\,r^{2\xi-2}\,.
\non
\qqq
We infer then that the velocity $v_r$ of size $r$ eddies 
scales as $\,\sim r^{\xi-1}$ corresponding to the eddy turnover 
time $\tau_r={r\over v_r}\sim r^{2-\xi}$. Within this limited 
comparison, $\xi={4\over 3}$ gives the Kolmogorov scaling 
of velocities. Of course, the Kraichnan ensemble, as Gaussian, 
is not intermittent. Also its odd-point correlation functions 
vanish. We shall see in next Lecture, however, that it
leads to intermittency in transport phenomena.
\vskip 0.3cm

In the velocity fields of the Kraichnan ensemble, we shall 
consider the Lagrangian trajectories, 
perturbed first by a small noise, i.e.\,\,satisfying the
stochastic differential equation:
\qq
{d\Nx\over ds}\ =\ \Nv(s,\Nx)\ +\ \sqrt{2\kappa}\,\m
{d{\bf\beta}(s)\over ds}\,,
\label{LTB}
\qqq
where ${\bf\beta}(s)$ is a $d$-dimensional Brownian motion
and $\kappa>0$. For a smooth velocity $\Nv$, solutions 
of Eq.\,\,(\ref{LTB}) form a Markov process which may be 
characterized by the transition probabilities
\qq
P^{t,s}(\Nx,\Ny\vert\Nv)\ 
=\ \overline{\delta(\Ny-\Nx_{t,\Nx}(s))}\,,
\label{LB}
\qqq
where $\Nx_{t,\Nx}(s)$ denotes the solution of the differential 
equation (\ref{LTB}) (with $s$ as the running time) 
passing at time $t$ by $\Nx$ and the overbar stands 
for the averaging with respect to the Brownian motion 
${\bf\beta}$. The transition probabilities satisfy 
the linear equation
\qq
(\da_t+\Nv(t,\Nx)\cdot\Nna_{\Nx}-\kappa\Nna_\Nx^2)\,\,
P^{t,s}(\Nx,\Ny\vert\Nv)\ =\ 0\,,
\non
\qqq
with the initial condition $\m P^{t,t}(\Nx,\Ny\vert\Nv)
=\delta(\Nx-\Ny)$, \m so that, in the operator language,
\qq
P^{t,s}(\Nv)\ =\ \CT\,\m\ee^{\,\int\limits_s^t
[-\Nv(\sigma)\cdot\Nna\,+\,\kappa\Nna^2]\, d\sigma}\m,
\non
\qqq
where $\CT$ denotes the time ordering
(we have assumed $t\geq s$ here, the case $t<s$ is treated 
similarly).
\vskip 0.3cm

As noticed by Y. Le Jan and O. Raimond, the transition
probabilities $P^{t,s}(\Nv)$ still make sense for velocities
$\Nv$ of the Kraichnan ensemble, despite the poor regularity 
properties of the latter. These authors have rewritten 
the above expression for $P^{t,s}(\Nv)$ in the Wick-ordered form:
\qq
P^{t,s}(\Nv)
&=&:\,\m\CT\,\m\ee^{\,\int\limits_s^t
[-\Nv(\sigma)\cdot\Nna\,+\,(\kappa+\hf D_0)\Nna^2]\, d\sigma}\,:\cr
&=&\sum\limits_{n=0}^\infty(-1)^n\int\limits_s^td\sigma_n\dots
\int\limits_s^{\sigma_2}d\sigma_1\,\,:\,\ee^{\m(\kappa+\hf D_0)
(t-\sigma_n)\Nna^2}\,\Nv(\sigma_n)\cdot\Nna\,\ee^{\m(\kappa+\hf D_0)
(\sigma_n-\sigma_{n-1})\Nna^2}\,
\dots\cr\cr
&&\hspace{3.2cm}\dots\,\,\ee^{\m(\kappa+\hf D_0)(\sigma_2-\sigma_1)
\Nna^2}\,
\Nv(\sigma_1)\cdot\Nna\,\m\ee^{\m(\kappa+\hf D_0)
(\sigma_1-s)\Nna^2}\,:\,,
\label{twor}
\qqq
which is an infinite sum of Wick ordered monomials
of $\Nv$ that may be represented by the diagrams of Fig.\,\,2. 

\leavevmode\epsffile[-190 -20 230 220]{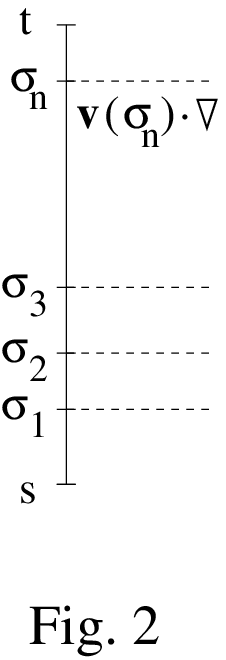}

The homogeneous Wick polynomials of different degrees are orthogonal 
in the $L^2$ scalar product w.r.t. the Gaussian measure 
of the velocity process. By establishing the bound 
\qq
\langle\,\vert P^{t,s}(\Nv)\m g\vert^2\rangle\ \leq\  
\ee^{\m\hf D_0(t-s)\m\Nna^2}\vert g\vert^2
\non
\qqq
where $g$ are functions on $\NR^d$, one shows then that 
the series giving $P^{s,t}(\Nv)\m f$ converges in the space 
of square-integrable functionals of $\Nv$, so also for almost 
all (a.a.) $\Nv$, \m as long as $g$ is bounded. It defines for a.a. 
velocities the transition probabilities $P^{t,s}(\Nx,\Ny\vert\Nv)$
of a Markov process which are continuous as functions 
of $\kappa\geq0$. Note that 
\qq
P_1^{t,s}(\Nx,\Ny)\ \equiv\ \langle\, P^{t,s}(\Nv)(\Nx,\Ny)\,\rangle\ 
=\ \ee^{\m(\kappa+\hf D_0)\vert t-s\vert\Nna^2}(\Nx,\Ny)\,.
\label{1pf}
\qqq
\vskip 0.3cm

The essential question that we want to address is about
the nature of the Markov process obtained in the limit
$\kappa\to0$. Are the limiting transition probabilities
concentrated at single points $\Ny$ leading to deterministic
Lagrangian trajectories determined, in a fixed velocity
realization, by the initial position or, on the contrary, 
do they stay diffuse? A way to study this question is to 
examine the joint PDF (probability distribution function) 
of the equal-time values of two solutions 
of Eq.\,\,(\ref{LTB}) averaged over the velocity ensemble:
\qq
P^{t,s}_2(\Nx_1,\Nx_2;\Ny_1,\Ny_2)\ =\ 
\langle\,P^{t,s}(\Nx_1,\Ny_1\vert\Nv)\,\,P^{t,s}(\Nx_2,
\Ny_2\vert\Nv)\,\rangle\,.
\non
\qqq
The average on the right hand side is given by the sum of terms
described by the diagrams of Fig.\,\,3 with the broken-line 
propagators corresponding to the spatial part $(D_0-d(\cdot))$ 
of the velocity 2-point functions (\ref{v2pt}). The whole sum 
becomes the perturbative expansion for the heat kernel of the 
second order differential operator:
\qq
P_2^{t,s}(\un{\Nx};\un{\Ny})\ 
=\ \ee^{-\vert t-s\vert\m M_2}(\un{\Nx};\un{\Ny})\,,
\label{HK}
\qqq
where
\qq
M_2&=&-(\kappa+\hf D_0)\,(\Nna_{\Nx_1}^2\,+\,
\Nna_{\Nx_2}^2)\ -(D_0\m\delta^{\alpha\beta}-d^{\alpha\beta}
(\Nx_1-\Nx_2))\nabla_{x_1^\alpha}\nabla_{x_2^\beta}\cr
&=&d^{\alpha\beta}(\Nx_1-\Nx_2))
\nabla_{x_1^\alpha}\nabla_{x_2^\beta}
\ -\ \kappa\,(\Nna_{\Nx_1}^2\,+\,\Nna_{\Nx_2}^2)
\ +\ \hf D_0\,
(\Nna_{\Nx_1}\,+\,\Nna_{\Nx_2})^2\,.
\label{MM2}
\qqq
The last term drops out in the translation-invariant sector.
In other words, two solutions of Eq.\,\,(\ref{LTB}) undergo,
in their relative motion, an effective diffusion with the
diffusion coefficient dependent on their relative position.

\leavevmode\epsffile[-160 -20 240 220]{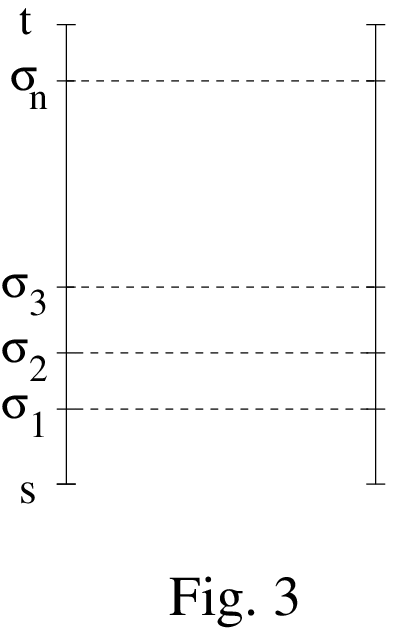}

The PDF  $\,P_2^{t,s}(r;\rho)\,$ of the distance $\rho$ 
between the time $s$ positions of two solutions, 
given their time $t$ distance $r$, is expressed by the heat kernel 
of the operator $M_2$ restricted to the translation and rotation 
invariant sector. The latter becomes an explicit second order 
differential operator in the radial variable:
\qq
M_2^{inv}\ =\ -C\, r^{\xi-a}\m\da_r\,r^a\m\da_r
\,-\,2\,\kappa\,r^{-d+1}\m\da_r\,r^{d-1}\m\da_r 
\label{sp}
\qqq
with the exponent $a=a(\xi,\wp)$ (for simplicity, we give 
the formula after the removal of the infrared cutoff $L$
in Eq.\,\,(\ref{msr})). The operator $M_2^{inv}$
may be transformed by a change of variables and a similarity
transformation to a Schr\"{o}dinger operator on the half-line.
In particular,
\qq
\lim\limits_{\kappa\to0}\ M_2^{inv}\ =\ C'\,
u^c\,\,[\,-\da_u^2\,+\,{_{b^2-{1\over 4}}\over^{u^2}}\,]\,\, u^{-c}\,,
\label{sop}
\qqq
where $u=r^{2-\xi\over 2}$ and $b=b(\xi,\wp)$, $c=c(\xi,\wp)$. 
The limit $\kappa\to0$ of the PDF $\,P_2^{t,s}(r,\rho)\,$ can 
be explicitly controled. Two different regimes appear in this limit.
\eject

\noindent 1. \ {\bf Weakly compressible regime} 
\vskip 0.3cm

For weak compressibility $\wp<{d\over\xi^2}$, which corresponds 
to $b<1$, the distance PDF $\,P_2^{t,s}(r,\rho)\,$ is an integrable 
function of $\rho$ and stays such in the limit $r\to0$:
\qq
\lim\limits_{r\to0}\ \lim\limits_{\kappa\to0}\ P_2^{t,s}(r;\rho)\,\,
d\rho\ \ \propto\ \ ({_{{\rho}^{2-\xi}}\over^{\vert t-s\vert}})^{1-b}\ 
\ee^{\m-\m{{\rho}^{2-\xi}\over 4C'\vert t-s\vert}}\,\,{_{d\rho}
\over^{\rho}}\,.
\label{nspd}
\qqq
This behavior excludes concentration of the transition 
probabilities $P^{t,s}(\Nx,\Ny\vert\Nv)$ at single points $\Ny$.
In particular, we obtain the Richardson dispersion law
in the form
\qq
\lim\limits_{r\to0}\ \lim\limits_{\kappa\to0}\ 
\int\rho^2\,\, P_2^{0,t}(r,\rho)\,\, d\rho\ \ \propto\ \ 
t^{^{2\over2-\xi}}
\non
\qqq
indicating an explosive separation of the Lagrangian trajectories 
and reproducing, for $\xi={4\over 3}$, the mean growth 
(\ref{prt}). As we see, the trajectories in a fixed 
typical realization of the velocity fields are not
determined by the initial position but rather form 
a Markov process with diffuse transition probabilities 
$\,P^{t,s}(\Nx,\Ny\vert\Nv)\m$, as predicted above,
see Fig.\,\,4.

\leavevmode\epsffile[-110 -25 270 220]{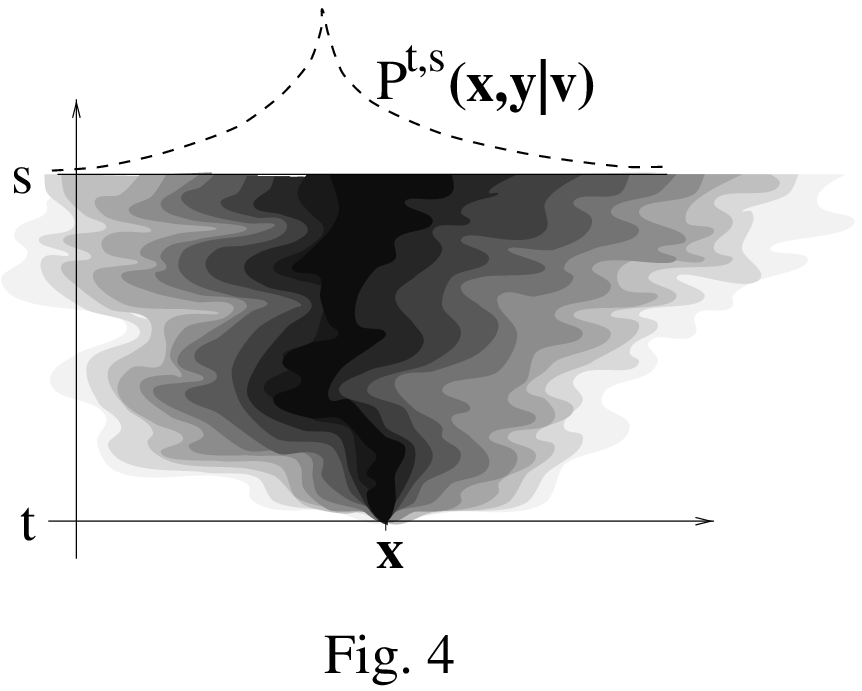}

\noindent 2. \ {\bf Strongly compressible regime} 
\vskip 0.3cm

For strong compressibility $\wp\geq{d\over\xi^2}$, i.e.\,\,for 
$b\geq1$, one observes a different behavior due to the strong 
repulsive singularity at $u=0$ in the operator (\ref{sop})\m:
\qq
\lim\limits_{\kappa\to0}\ P_2^{t,s}(r;\rho)\,\,
d\rho\ =\ p^{t,s}(r)\,\m\delta(\rho)\,\m d\rho\ +\ {\rm regular}
\non
\qqq
with the coefficient $p^{t,s}(r)$ of the delta-function 
converging to $1$ when $r\to0$ or $|t-s|\to\infty$. In particular,
\qq
\lim\limits_{r\to0}\ \lim\limits_{\kappa\to0}\ P_2^{t,s}(r;\rho)\,\, 
d\rho\ =\ \delta(\rho)\,\m d\rho
\non
\qqq
in this regime implying the concentration of the transition 
probabilities $P^{t,s}(\Nx,\Ny\vert\Nv)$ at single points $\Ny$ 
and the existence, in a fixed typical velocity realization, 
of Lagrangian trajectories determined by their initial positions. 
The presence of the singular term in 
$\,P_2^{t,s}(r;\rho)\,$ for $\kappa=0$ indicates, however,
an {\bf implossive collapse} of distinct Lagrangian 
trajectories with a positive probability which grows in time,
see Fig.\,\,5.
\vskip 0.3cm

We infer that in H\"{o}lder-continuous velocity fields, 
there is a competition of the tendency of two Lagrangian trajectories
to separate or to collapse explosively. For weak compressibility,
this is the first tendency that wins. The trapping
effects increase, however, with the degree of compressibly 
and lead, in the Kraichnan ensemble, to a sharp transition 
in the behavior of the trajectories at $\wp={d\over\xi^2}$.

\leavevmode\epsffile[-130 -30 270 220]{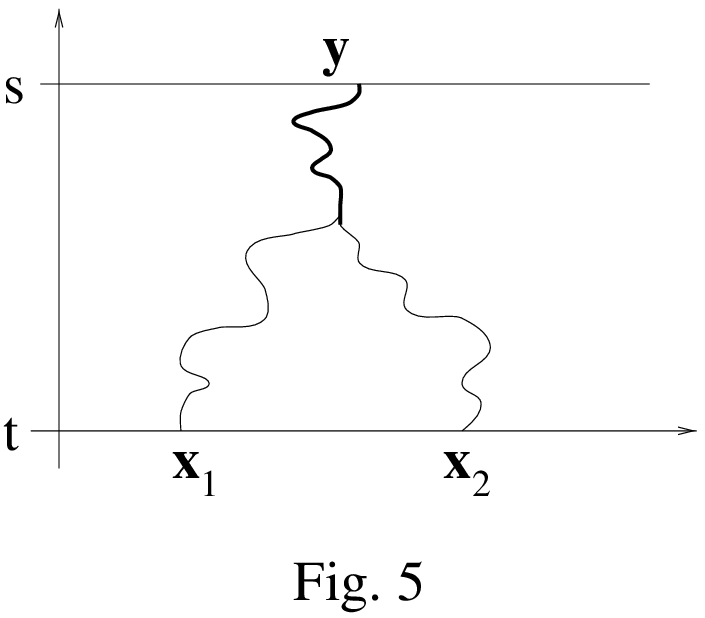}
\vskip 0.3cm

\noindent{\bf Bibliography}
\vskip 0.3cm

\noindent The Richardson dispersion law was formulated in:
\vskip 0.1cm

1. \ L. F. Richardson: {\it Atmospheric diffusion shown on 
a distance-neighbour graph}. Proc. R. Soc. Lond. {\bf A 110} (1926),
709-737
\vskip 0.3cm

\noindent Weak solutions of Euler equations with
Lagrangian trajectories forming a stochastic process
were discussed in:
\vskip 0.1cm

2. \ Y. Brenier: {\it A homogenized model for vortex sheet}.
Arch. Rat. Mech. Anal. {\bf 138} (1997), 319-363
\vskip 0.1cm

3. \ A. Shnirelman: {\it Weak solutions with decreasing energy
of incompressible Euler equations}. IHES/M/99/02 preprint
\vskip 0.3cm

\noindent The introduction of velocities decorrelated in time 
in turbulence studies goes back to:
\vskip 0.1cm

4. \ R. H. Kraichnan: {\it Small-scale structure of a scalar
field convected by turbulence}.
Phys. Fluids {\bf 11} (1968), 945-963
\vskip 0.3cm

\noindent The construction of the transition probabilities
$\,P^{t,s}(\Nx,\Ny\vert\Nv)\,$ was described in:
\vskip 0.1cm

5. \ Y. Le Jan and O. Raimond: {\it Solution statistiques fortes
des \'{e}quations diff\'{e}rentielles stochastiques}. 1998 Orsay
preprint
\vskip 0.3cm

\noindent The two regimes of behavior of the Lagrangian 
trajectories in the Kraichnan ensemble of velocities were
analyzed in:
\vskip 0.1cm

6. \ K. Gaw\c{e}dzki and M. Vergassola: {\it Phase transition
in the passive scalar advection}. cond-mat/9811399, to appear 
in Physica D
\vskip 2cm
\eject
\ \vskip 1cm

\no {\large\bf{LECTURE 4}}
\vskip 0.5cm

What are the consequences on the hydrodynamical properties 
of the flows of the observed dramatic behaviors of Lagrangian 
trajectories, violating the Newton-Leibniz paradigm about 
existence and uniqueness of solutions of the ODE's? The answer 
to this question may be the clue to a consistent theory 
of developed turbulence. Here, we shall content ourselves 
with discussing the transport properties in the flows induced 
by the velocities of the Kraichnan ensemble. We shall see 
that these properties differ drastically for weak and strong
compressibility as a result of different behaviors 
of the Lagrangian trajectories. For concreteness, we shall 
look at the passive transport of the scalar quantity 
$\theta(t,\Nx)$ (called tracer) whose evolution is described
by the advection-diffusion equation
\qq
\da_t\theta\s+\s(\Nv\cdot\Nna)\theta\s-
\s\kappa\Nna^2\theta\s=\s f\,,
\label{ps}
\qqq
where $f(t,\Nx)$ denotes now a scalar source. 
In the incompressible case, $\theta$ may also describe 
the temperature field or the density of pollutant. 
The passivity of the advection means that the back reaction 
of $\m\theta$ on the velocity is ignored. 
\vskip 0.3cm

When the velocity field is sufficiently smooth, it is easy 
to solve the above linear equation for $\theta$. 
For $\kappa=0$ and $f=0$, the scalar is simply carried by
the Lagrangian flow
\qq
\theta(t,\Nx)\ =\ \theta(s,\Nx_{t,\Nx}(s))\,,
\non
\qqq
where $\Nx_{t,\Nx}(s)$ is the Lagrangian trajectory passing
at time $t$ by point $\Nx$. Note that the forward evolution 
of $\m\theta$ corresponds to the backward Lagrangian flow.
In the presence of the source $f$, the scalar
is also created or depleted along the trajectory:
\qq
\theta(t,\Nx)\ =\ \theta(s,\Nx_{t,\Nx}(s))\,+\,\int\limits_s^t
f(\sigma,\Nx_{t,\Nx}(\sigma))\,\m d\sigma\,.
\non
\qqq
Finally, when $\kappa\not=0$, $\,\Nx_{t,\Nx}(s)$ 
should be taken as a solution of the equation (\ref{LTB}) 
for the Lagrangian trajectories perturbed by the Brownian
motion and the above formulae should be averaged over 
the latter. Thus
\qq
\theta(t,\Nx)&=&\int\ov{\delta(\Ny-\Nx_{t,\Nx}(s))}\,\m\theta(s,\Ny)
\,\m d\Ny\ +\ \int\limits_s^t\Big(\int\ov{\delta(\Ny-\Nx_{t,\Nx}(\sigma))}
\,\m f(\sigma,\Ny)\,\,d\Ny\Big)\m d\sigma\cr
&=&\int P^{t,s}(\Nx,\Ny\vert\Nv)\,\m\theta(s,\Ny)
\,\m d\Ny\ +\ \int\limits_s^t\Big(\int P^{t,\sigma}(\Nx,\Ny\vert\Nv)
\,\m f(\sigma,\Ny)\,\,d\Ny\Big)\m d\sigma\,,
\label{zap}
\qqq
see Eq.\,\,(\ref{LB}). The right hand side still makes sense for 
a.a. velocities of the Kraichnan ensemble and defines
a weak solution of the linear differential equation (\ref{ps}), i.e.
the one satisfying the equation in the distributional 
sense\footnote{Eq.\,\,(\ref{ps}) is an infinite-dimensional 
stochastic differential equation due to the white temporal dependence 
of $\Nv$ and it should be treated according to the Stratonovich
prescription; the weakness of the solution is, however,  
due to its poor spatial regularity resulting from the 
non-differentiability in space of typical velocities 
and the related stochastic character of the Lagrangian 
trajectories}.
\vskip 0.3cm

First, let us assume that we are given a random distribution
of the scalar at time $s$, independent of the (later) velocities, 
and we wish to study its distribution at the later time $t$.
In free decay, i.e.\,\,in the absence of the source $f$,
by taking averages over the initial distribution and over 
the velocities, we obtain for the 1-point function of $\m\theta\m$:
\qq
\langle\,\theta(t,\Nx)\m\rangle\, 
=\,\int P_1^{t,s}(\Nx,\Ny)\,\,\langle\,\theta(s,\Ny)\m\rangle
\,\,d\Ny\,=\,\int\ee^{-(s-t)(\kappa+\hf D_0)\Nna^2}(\Nx,\Ny)\,\,
\langle\,\theta(s,\Ny)\m\rangle\,\,d\Ny\,,\hspace{0.3cm}
\non
\qqq
see Eq.\,\,(\ref{1pf}).
In other words, the $1$-point function decays diffusively. 
Note that the initial diffusivity $\kappa$ is increased
by the {\bf eddy diffusivity} $\hf D_0$.
Similarly, for the 2-point function,
\qq
\langle\,\theta(t,\Nx_1)\,\theta(t,\Nx_2)\m\rangle
\ =\ \int P_2^{t,s}(\Nx_1,\Nx_2;\Ny_1,\Ny_2)\,\,
\langle\,\theta(s,\Ny_1)\,\theta(s,\Ny_2)\m\rangle
\,\,d\Ny_1\, d\Ny_2\,
\label{2p1}
\qqq
and, for the $N$-point one,
\qq
\langle\,\prod\limits_{n=1}^N\theta(t,\Nx_n)\m\rangle
\ =\ \int P_{_N}^{t,s}(\un{\Nx};\un{\Ny})\,\,
\langle\,\prod\limits_{n=1}^N\theta(s,\Ny_n)\m\rangle\,\,d\un{\Ny}\,,
\non
\qqq
where
\qq
P_{_N}^{t,s}(\un{\Nx};\un{\Ny})\ =\ \langle\,\prod\limits_{n=1}^N
P^{t,s}(\Nx_n,\Ny_n\vert\Nv)\,\rangle\ =\ P_{_N}^{s,t}(\un{\Nx};\un{\Ny})
\non
\qqq
is the joint PDF of the time $s$ positions $\un{\Ny}$ 
of the Lagrangian trajectories, given their time $t$ positions 
$\un{\Nx}$, see Fig.\,\,6 (the last equality follows from the 
stationarity and time-reflection invariance of the Kraichnan ensemble).

\leavevmode\epsffile[-95 -20 295 205]{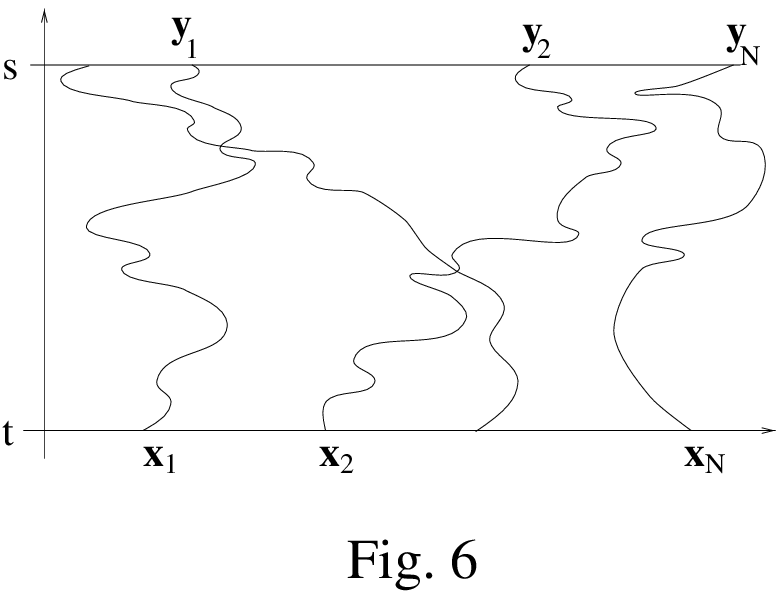}

\noindent It is easy to see that the PDF's $P_{_N}^{t,s}
(\un{\Nx};\un{\Ny})$ are again given by the heat kernels 
of second order differential operators:
\qq
P_{_N}^{t,s}(\un{\Nx};\un{\Ny})\ 
=\ \ee^{-\vert t-s\vert\m M_{_N}}(\un{\Nx}
;\un{\Ny})\,,
\non
\qqq
where
\qq
M_{_N}\ =\ \sum\limits_{n<m}d^{\alpha\beta}(\Nx_n-\Nx_m))
\nabla_{x_n^\alpha}\nabla_{x_m^\beta}\ 
-\ \kappa\sum\limits_{n=1}^N\Nna_{\Nx_n}^2\ +\ \hf D_0\,
(\sum\limits_n\Nna_{\Nx_n})^2\,.
\label{MMN}
\non
\qqq
\vskip 0.3cm

Assuming a homogeneous and isotropic initial distribution of 
$\theta$, we may replace the PDF $P_2^{t,s}(\Nx_1,\Nx_2;\Ny_1,\Ny_2)$ 
in Eq.\,\,(\ref{2p1}) by its translation- and rotation-invariant 
version $P_2^{t,s}(r,\rho)$. In particular, taking the limit $r\to0$, 
we infer that, for $\kappa=0$ and in the weakly compressible case 
$\wp<{d\over\xi^2}$, the mean scalar energy density
\qq
{\bar e}_\theta(t)\ \equiv\ \langle\m\hf\m\theta(t)^2\rangle
\ =\ \hf\int P_2^{t,s}(0,\rho)\,\,
\langle\,\theta(s,\Ny)\,\theta(s,{\bf 0})\m\rangle\,\,d\rho
\ <\ {\bar e}_\theta(s)
\label{inq}
\qqq
if the 2-point function 
$\m\langle\,\theta(s,\Ny)\,\theta(s,{\bf 0})\m\rangle\m$
decays for large $\rho=\vert\Ny\vert$. 
Indeed, the latter is bounded by its value 
at $\rho=0$ so that the relation (\ref{inq}) follows since 
$P_2^{t,s}(0,\rho)$ is a strictly positive probability density, 
see Eq.\,\,(\ref{nspd}). Hence the mean energy density 
of the scalar decreases with time (i.e.\,\,is dissipated) even 
for $\kappa=0$. On the contrary, in the strongly compressible case 
$\wp\geq{d\over\xi^2}$, the limit PDF $P_2^{t,s}(0,\rho)=\delta
(\rho)$ and the mean energy density is conserved in the unforced 
evolution at $\kappa=0\m$:
\qq
{\bar e}_\theta(t)\ =\ {\bar e}_\theta(s)\,.
\non
\qqq
\vskip 0.3cm

In the presence of the source $f$ which, as before, we shall assume
random Gaussian, independent of velocities and the initial
distribution of $\m\theta$, with mean zero and 
covariance
\qq
\langle\m f(t,\Nx)\s f(s,\Ny)\rangle\s=\s\delta(t-s)\s\s 
\chi({_{\vert\Nx-\Ny\vert}\over^L})\,,
\label{rf2}
\qqq
the 1-point function of the scalar diffuses as before
and for the 2-point function, we obtain from Eq.\,\,(\ref{zap})
\qq
\langle\,\theta(t,\Nx_1)\,\theta(t,\Nx_2)\m\rangle
=\int P^{t,s}_2(\un{\Nx};\un{\Ny})\, 
\langle\,\theta(s,\Ny_1)\,\theta(s,\Ny_2)\m\rangle\,\m
d\un{\Ny}+\int\limits_s^t\hspace{-0.06cm}
\Big(\int P_2^{t,\sigma}(\un{\Nx};
\un{\Ny})\,\chi({_{\vert\Ny_1-\Ny_2\vert}\over^L})\,\m 
d\un{\Ny}\Big)d\sigma.\hspace{0.5cm}
\label{2pff}
\qqq
The 2-point function solves now the equation
\qq
&&\da_t\,\langle\,\theta(\Nx_1)\,\theta(\Nx_2)\m\rangle\ =\ 
-M_2\,\m\langle\,\theta(\Nx_1)\,\theta(\Nx_2)\m\rangle\ +
\ \chi({_{\vert\Nx_1-\Nx_2\vert}\over^L})\cr\cr
&&=\ -\m d^{\alpha\beta}(\Nx_1-\Nx_2)\m\,
\langle\m\nabla_{\alpha}\theta(\Nx_1)\,
\nabla_\beta\theta(\Nx_2)\m\rangle
\ -\ 2\m\kappa\,\m\langle\m\Nna\theta(\Nx_1)\cdot\Nna\theta(\Nx_2)
\m\rangle
\ +\ \chi({_{\vert\Nx_1-\Nx_2\vert}\over^L})\,,
\label{2pfg}
\qqq
see Eqs.\,\,(\ref{HK}) and (\ref{MM2}). This is the passive scalar
counterpart of the NS relation (\ref{rel0}) that we discussed
in Lecture 2. In the weakly compressible regime $\wp<{d\over\xi^2}$,
the scalar 2-point function reaches a stationary state.
Arguing as before for the NS case, we infer in this state
the energy balance
\qq
{\bar\epsilon}_\theta\ =\ \hf\m\chi(0)
\non
\qqq
for the scalar mean dissipation rate ${\bar\epsilon}_\theta
=\langle\m\kappa(\Nna\theta)^2\rangle$ and the relation
\qq
\hf\m d^{\alpha\beta}(\Nx_1-\Nx_2)\s\nabla_{x_1^\alpha}
\nabla_{x_2^\beta}\s\,\langle\m\theta(\Nx_1)\s\theta(\Nx_2)\m\rangle\s
=\s\hf\m\chi({_{\vert\Nx_1-\Nx_2\vert}\over^L})\,.
\non
\qqq
The latter may be easily solved for the stationary 2-point function 
of the scalar giving
\qq
\langle\m \theta(\Nx)\s \theta({\bf 0})\m\rangle\s=\s A_2(\chi)
\s L^{2-\xi}\s-\s
{\rm const}.\s\s{\bar\epsilon}_\theta\s\m r^{2-\xi}\s
+\s\CO(L^{-2})\,,
\label{kr2}
\qqq
where $r\equiv|\Nx|$ or, for the scalar 2-point structure function,
\qq
S_2(r)\,\equiv\,\langle\m(\theta(\Nx)-\theta(0))^2\rangle\ \s\propto
\ \s{\bar\epsilon}_\theta
\s\m r^{2-\xi}
\label{kr3}
\qqq
for $\kappa=0$ and $r\ll L$. This is an analogue of the
Kolmogorov \m${4\over 5}\m$ law. It may be strengthen
to the operator product expansion for the $\kappa\to0$
limit of the dissipation operator \s$\epsilon_\theta
=\kappa(\Nna\theta)^2\s$
\qq
\epsilon_\theta(\Nx)\ =\ 
\lim\limits_{\Ny\to\Nx}\s\s
d^{\alpha\beta}(\Nx-\Ny)\s\s\nabla_{\alpha}\theta(\Nx)\s\m
\nabla_{\beta}\theta(\Ny)\s\bigg\vert_{{\kappa=0}}
\label{das}
\qqq
valid inside expectations in the limit \s$\kappa\to0\m$
and also in fixed realizations of $\theta$.
Eq.\s\s(\ref{das}) expresses the dissipative anomaly
in the weakly compressible regime of the Kraichnan model, 
analogous to the dissipative anomaly (\ref{da}) 
for the Navier-Stokes case. One may also check directly
the approximate constancy of the scalar energy flux 
towards large wavenumbers, establishing the existence 
of a direct scalar energy cascade.
\vskip 0.3cm

In the strongly compressible regime $\wp>{d\over\xi^2}$,
the behavior of the scalar 2-point function (\ref{2pff}) in the limit
$\kappa\to0$ is quite different. Now the 2-point function does 
not stabilize but has a constant contribution growing linearly in time.
The dissipation rate vanishes and scalar energy is pumped 
into the constant mode at a constant rate equal to the injection 
rate $\hf\m\chi(0)$. This signals the presence 
of an inverse cascade of scalar energy towards small wavenumbers.
The 2-point structure function of the scalar stabilizes, however,
and its stationary limit satisfies the equation
\qq
M_2\,\,\langle\m(\theta(\Nx)-\theta({\bf0})^2\m\rangle\ =\ 
2\m(\chi(0)-\chi({_{|\Nx|}\over^L}))
\non
\qqq
from which one infers that 
\qq
S_2(r)\ \propto\ r^{2-\xi}
\non
\qqq
for $r\gg L$.
\vskip 0.3cm

As for intermittency of the scalar statistics, 
the two regimes also show very different behaviors.
The question here is whether the higher structure
functions of the scalar \s$S_{_N}(r)\equiv\langle
(\theta(\Nx)-\theta(0))^{N}
\rangle\s$ scale with powers \s${N\over2}(2-\xi)\s$
as the dimensional analysis would suggest, in analogy
to the Kolmogorov theory. Although, by assumption, 
in the Kraichnan model there is no intermittency 
in the statistics of the velocity differences, 
the numerical studies of the incompressible
model indicate strong intermittency of the scalar differences 
signaled by anomalous values of the scaling exponents. 
Unlike in the NS case, we have now some analytic understanding
of this phenomenon. 
\vskip 0.3cm

First, it is not difficult to see that the higher equal-time
correlators of the scalar satisfy the evolution equations 
generalizing Eq.\,\,(\ref{2pfg})\m:
\qq
\da_t\,\m\langle\m\prod\limits_{n=1}^N\theta(\Nx_n)\,\rangle
\ =\ -\m M_{_N}\,\,\langle\m\prod\limits_{n=1}^N\theta(\Nx_n)\,\rangle
\ +\ \sum\limits_{p<q}\,\langle\prod\limits_{n\not=p,q}
\theta(\Nx_n)\,\rangle\,\,\chi({_{|\Nx_p-\Nx_q|}\over^L})\,.
\label{evN}
\qqq
In the weakly compressible regime, the
correlation functions stabilize for long time
and, besides, we expect that the limits $t\to\infty$
and $\kappa\to0$ commute. The stationary
equal time correlations satisfy a similar equation
but with the vanishing left hand side. By inverting operators
$M_{_N}$, we may then compute the stationary $N$-point functions 
of the scalar recursively, a rare situation, indeed, since
in most hydrodynamical problems the evolution equations
for the correlation functions, called the Hopf equations, do not 
close. A semi-rigorous analysis shows that, for small 
$\xi$ in the limit $\kappa\to0$ and on short distances 
or for large injection scale $L$,
\qq
\langle\m\prod\limits_{n=1}^N\theta(\Nx_n)\,\rangle
\ =\ A_{_N}(\chi)\,\,L^{{N\over2}(2-\xi)-\zeta_{_N}}
\,\,\varphi_{_N}(\un{\Nx})\ +\ 
\CO(L^{-2+\CO(\xi)})\ +\ ...\,,
\non
\qqq

\input{psfig}

\begin{figure}
\centerline{\psfig{file=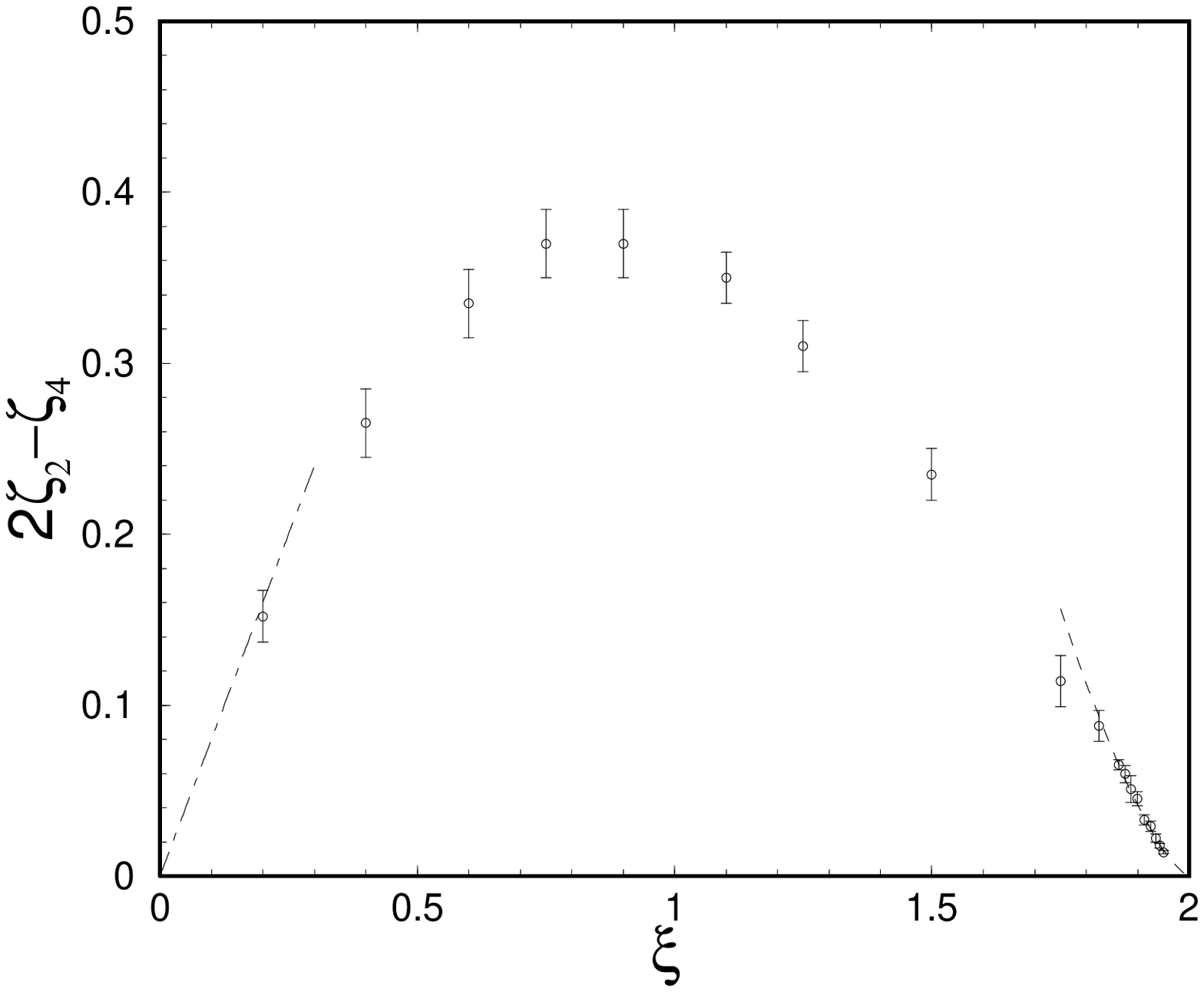,width=7.5cm,clip=}}
\end{figure}

\centerline{\Large{Fig. 7}}
\vskip 1cm

\noindent where $\varphi_{_N}(\un{\Nx})$ are scaling zero modes
of the operators $M_{_N}$, i.e.
\qq
M_{_N}\,\m\varphi_{_N}(\un{\Nx})\ =\ 0\,,\qquad\varphi_{_N}
(\lambda\un{\Nx})\,=\,\lambda^{\zeta_{_N}}\,\m\varphi_{_N}
(\un{\Nx})
\non
\qqq
with
\qq
\zeta_{_N}\ =\ {_N\over^2}(2-\xi)\,-\,{_{N(N-2)(1+2\wp)}
\over^{2(d+2)}}\,\xi\ +\ \CO(\xi^2)\,.
\non
\qqq
The coefficients $A_{_N}(\chi)$ are non-universal amplitudes
and the dots denote terms that do not depend of, at least, 
one variable and, as such, do not contribute 
to the correlations of scalar differences.
In particular, the structure functions
\qq
S_{_N}(r)\ \propto\ L^{{N\over2}(2-\xi)
-\zeta_{_N}}\,\,r^{\zeta_{_N}}\,.
\non
\qqq
This {\bf zero mode dominance} of the stationary higher-point
functions of the scalar (note that such zero modes 
drop out in the stationary version of Eq.\,\,(\ref{evN}))
has been exhibited by the perturbative analysis of the Green
functions of operators $M_{_N}$ around $\xi=0$.
It has been confirmed by perturbative analyses in powers 
of the inverse dimension and of $(2-\xi)$ and by numerical results, 
see Fig.\,\,7 representing the values of the simulations 
by Frisch-Mazzino-Vergassola of the 4-point function anomalous 
exponent in the three-dimensional incompressible Kraichnan model.
\vskip 0.3cm

What is the physical meaning of the zero modes of the operators
$M_{_N}$ that dominate the short-distance asymptotics of
the $N$-point functions of scalar differences? They are
{\bf slow modes} of the effective diffusion
of Lagrangian trajectories with generators $M_{_N}$. Indeed, 
for generic scaling function $\psi_{_N}(\un{\Nx})$ of scaling
dimension $\sigma_{_N}$, viewed as a function of time
$t$ positions of the Lagrangian trajectories, the effective 
time evolution is
\qq
\langle\m\psi_{_N}\m\rangle_{_t}\ \equiv\ \int
P_{_N}^{0,t}(\un{\Nx},\un{\Ny})\,\,\psi_{_N}(\un{\Ny})\,\, d\Ny\ \ 
\,\sim\ \ \,t^{\sigma_{_N}\over2-\xi}
\label{tgr}
\qqq
for large $t$, i.e.\,\,it exhibits a {\bf super-diffusive} growth.
But if $\psi_{_N}=\phi_{_N}$ is a zero mode of $M_{_N}$ then
the above expectation is conserved in time (such statistically
conserved modes are accompanied by descendent ones for which
the time growth is slower than (\ref{tgr})).
\vskip 0.3cm

In the strongly compressible phase with the inverse cascade
of scalar energy, the behavior of the higher structure
functions is different. In fact, only the lower ones
stabilize, but the ones that do, scale normally on large 
distances. In this regime one can find exactly 
the stationary form of the PDF of the scalar difference:
\qq
\langle\s\delta(\vartheta\,-\,{_{\theta(\Nx)\,-\,\theta({\bf 0})}\over^
{r^{(2-\xi)/2}}})\,\rangle
\ \ \propto\ \ [\chi(0)\,+\, C'
\,\vartheta^2]^{-b-\hf}
\non
\qqq
for $r\gg L$. Its scaling form indicates that there is 
no intermittency in the inverse cascade of the scalar
(the non-Gaussianity is scale-independent). Its poor
decay at infinity corresponds to the fact that only
lower structure functions reach a stationary regime.
\vskip 0.5cm

As we see, the transport of a scalar tracer by velocities
distributed according to the Kraichnan ensemble shows two
different phases characterized by different direction of
the scalar energy cascades and different degrees of intermittency.
The phase transition occurs at the value $\wp={d\over\xi^2}$
of the compressibility degree, where the behavior of the
Lagrangian trajectories changes drastically from 
the explosive separation to the implosive collapse. These two
phases are quite reminiscent of the behavior of the three
dimensional versus two-dimensional developed turbulence.
That suggests that one should put more stress 
on the Lagrangian methods in studying the latter,
especially on the properties of the Lagrangian flow 
in the weak solutions of the Euler equation. 
Of course the NS or the Euler equation, unlike the scalar advection 
one, are non-linear. They describe velocity fields that are not 
only carried by their own Lagrangian trajectories but also 
stretched and there are non-local effects due 
to pressure. Some of these effects, however, may 
be studied already in various models of passive 
advections (passive vectors, linearized NS equation, etc.). 
It seems that the study of easy models of turbulence 
has a potential to teach us important lessons that we have 
to master to stand a chance of solving the 
fully-fledged problem of developed turbulence.
\vskip 0.6cm

\noindent{\bf Bibliography}
\vskip 0.3cm

\noindent For the early works on passive scalar
advection see:
\vskip 0.1cm

1. \ A. M. Obukhov: {\it Structure of the temperature
field in a turbulent flow}. Izv. Akad. Nauk SSSR, Geogr. 
Geofiz. {\bf 13} (1949), 58-69
\vskip 0.1cm

2. \ S. Corrsin: {\it On the spectrum of isotropic temperature
fluctuations in an isotropic turbulence}. J. Appl. Phys. 
{\bf 22} (1951), 469-473
\vskip 0.3cm

\noindent That the scalar 2-point function may be found
exactly if velocities are time decorrelated was realized
in Kraichnan's 1968 paper cited at the end of previous Lecture.
\vskip 0.3cm

\noindent The presence of scalar intermittency in the
incompressible Kraichnan model model was first suggested in:
\vskip 0.1cm

3. \ R. H. Kraichnan: {\it Anomalous scaling of a randomly advected
passive scalar}. Phys. Rev. Lett. {\bf 72} (1994), 1016-1019
\vskip 0.3cm

\noindent The zero-mode dominance picture of the scalar intermittency
has been proposed in
\vskip 0.1cm

4. \ B. Shraiman and E. Siggia:
{\it Anomalous scaling of a passive scalar
in turbulent flow}. C.R. Acad. Sci. {\bf 321} (1995), 279-284
\vskip 0.1cm

5. \ K. Gaw\c{e}dzki and A. Kupiainen:
{\it Anomalous scaling of the passive scalar}.
Phys. Rev. Lett. {\bf 75} (1995), 3834-3837
\vskip 0.1cm

6. \  M. Chertkov, G. Falkovich, I. Kolokolov and V. Lebedev: 
{\it Normal and anomalous acaling
of the fourth-order correlation function of a randomly
advected scalar}. Phys. Rev. {\bf E 52} (1995), 4924-4941
\vskip 0.3cm

\noindent The numerical data of Fig.\,\,7 come from:
\vskip 0.1cm

7. \ U. Frisch, A. Mazzino and M. Vergassola: 
{\it Intermittency in Passive Scalar Advection}. 
Phys. Rev. Lett. {\bf 80} (1998), 5532-5535
\vskip 0.3cm

\noindent The phase transition with increase of compressibility
in the $\xi=2$ limiting case of the Kraichnan model was
observed in:
\vskip 0.1cm

8. \ M.~Chertkov, I.~Kolokolov and M.~Vergassola: {\it Inverse 
versus direct cascades in turbulent advection},  Phys. Rev. Lett. 
{\bf 80}, 512-515 (1998)
\vskip 0.3cm

\noindent The analysis of the phase transition for $\xi<2$
was done in the 1988 preprint of M. Vergassola and the present author
cited at the end of previous Lecture.
\vskip 0.3cm

\noindent For some more details on the Kraichnan model discussed
in an informal way, see:
\vskip 0.1cm

9. \ K.~Gaw\c{e}dzki: {\it Intermittency of Passive Advection}, 
in ``Advances in Turbulence VII'', U.~Frisch ed. Kluwer Acad. Publ. 
1998, pp. 493-502

\end{document}